\documentclass[aps,prx,superscriptaddress,twocolumn, secnumarabic, amssymb, floatfix, 10pt]{revtex4-2}

\usepackage[T1]{fontenc}
\usepackage[nolist]{acronym}
\usepackage{microtype}
\usepackage{graphicx} 
\usepackage{lipsum} 
\usepackage[english]{babel}
\usepackage{amsmath}
\usepackage{hyperref}
\usepackage{cancel}
\usepackage{bm}
\usepackage{multirow}
\usepackage{colortbl,hhline}
\usepackage{siunitx}
\usepackage{color}
\usepackage[percent]{overpic}

\usepackage{blindtext}
\usepackage{xr}

\newcommand{\ti}{\tau_i}
\newcommand{\td}{\tau_{D}}

\newcommand{\tm}{\tau_{m}}
\newcommand{\tdi}{\tau_{D,i}}
\newcommand{\tmi}{\tau_{m,i}}

\newcommand{\tmfp}{\tau_{\mathrm{MFP}}}

\newcommand{\ddx}{\partial_x}

\newcommand{\noiseLoc}{\eta[x(t),t]}

\newcommand{\xDL}{\tilde{x}}
\newcommand{\yDL}{\tilde{y}}

\newcommand{\tDL}{\tilde{t}}
\newcommand{\GammaDL}{\widetilde \Gamma}
\newcommand{\FDL}{\widetilde F}
\newcommand{\XDL}{\tilde{X}}

\newcommand{\ddxDL}{\partial_{\tilde x}}

\newcommand{\fDL}{\tilde{f}}
\newcommand{\UDL}{\tilde{U}}

\newcommand{\noiseLocDL}{\tilde{\eta}[\tilde{x}(\tilde{t}),\tilde{t}]}

\newcommand{\GW}{\Gamma_{\mathrm{W}}}
\newcommand{\GB}{\Gamma_{\mathrm{B}}}

\newcommand{\gW}{\gamma_{\mathrm{W}}}
\newcommand{\gB}{\gamma_{\mathrm{B}}}

\newcommand{\tmW}{\tau_{m,\mathrm{W}}}
\newcommand{\tW}{\tau_{\mathrm{W}}}
\newcommand{\tB}{\tau_{\mathrm{B}}}

\newcommand{\tdW}{\tau_{D,\mathrm{W}}}
\newcommand{\tdB}{\tau_{D,\mathrm{B}}}

\newcommand{\OmegaMax}{\omega_{\mathrm{max}}}
\newcommand{\OmegaMin}{\omega_{\mathrm{min}}}

\newcommand{\vthres}{v_{t}}

\graphicspath{{./figs/}}

\begin{document}

\title{Barrier-crossing times for different non-Markovian friction\\ in well and barrier -- A numerical study}

\author{Florian N. Br\"unig}
\affiliation{Department of Physics, Freie Universit\"at Berlin, 14195 Berlin, Germany}

\author{Roland R. Netz}
\affiliation{Department of Physics, Freie Universit\"at Berlin, 14195 Berlin, Germany}

\author{Julian Kappler}
\affiliation{Department of Applied Mathematics and Theoretical Physics, Centre for Mathematical Sciences, University of Cambridge, Wilberforce Road, Cambridge CB3 0WA, United Kingdom}

\date{\today}
\begin{abstract}
We introduce a generalized Langevin model system for different non-Markovian effects in the well and barrier regions of a potential, and use it to numerically study 
the dependence of the barrier-crossing time.
In the appropriate 
limits, our
 model interpolates between the theoretical barrier-crossing-time predictions 
 by Grote and Hynes (GH),  as well as by Pollak \textit{et al.},
  which for a single barrier memory time can differ by several orders of magnitude.
Our model furthermore allows to test an analytic rate theory for space-inhomogeneous memory, which disagrees with our numerical results in the long well-memory regime.
In this regime, we find that short barrier memory decreases the barrier-crossing time as compared to long
barrier memory.
This is in contrast with the short well-memory regime, where both our numerical results and GH theory
predict an acceleration of the barrier crossing time with increasing barrier memory time.
Both effects, the `Markovian-barrier acceleration' and GH `non-Markovian-barrier acceleration' can be understood from a committor analysis.
Our model combines finite relaxation times of orthogonal degrees of freedom with a space-inhomogeneous coupling to such degrees, and represents a step towards more realistic modeling of physical reaction coordinates.
\end{abstract}


\maketitle

\section{Introduction}

Many physical systems are comprised of large numbers of interacting degrees of freedom. 
A standard approach towards understanding dynamics in such systems is to define a low-dimensional reaction coordinate, motivated by the phenomenon to be investigated, 
and to construct an effective model for the dynamics of this reaction coordinate \cite{Zwanzig1961,
Mori1965a,
Zwanzig1973,
Zwanzig2001,
Berezhkovskii2005,
Lange2006,
Daldrop2018,
Kappler2019b,
Satija2019,
Lickert2020}.
Hereby, the orthogonal degrees of freedom are subsumed into an effective
heat bath, which interacts with the reaction coordinate \cite{Zwanzig1961,
Mori1965a,
Zwanzig1973,Zwanzig2001}.
One is then typically interested in the long-time dynamics of the reaction coordinate and in particular rare events such as barrier-crossing phenomena characterized by \acp{MFPT} 
 \cite{Kramers1940,
Chandler1978,
Grote1980,
Chandler1986,
Melnikov1986,
Hanggi1990,
Melnikov1991,
Best2006,
Daldrop2018}.
Systems where this approach has been applied are molecules in solution, which show conformational transitions, for example protein-folding \cite{Wilemski1974,
Szabo1980,
Daldrop2018,
Kappler2019,Ayaz2021},
or chemical reactions, where the reaction coordinate characterizes the 
transition from reactants to products \cite{Grote1980, Straub1987, Ciccotti1990, Benjamin1991, Rey1992a, Annapureddy2014, Meyer2021}.

\begin{figure*}
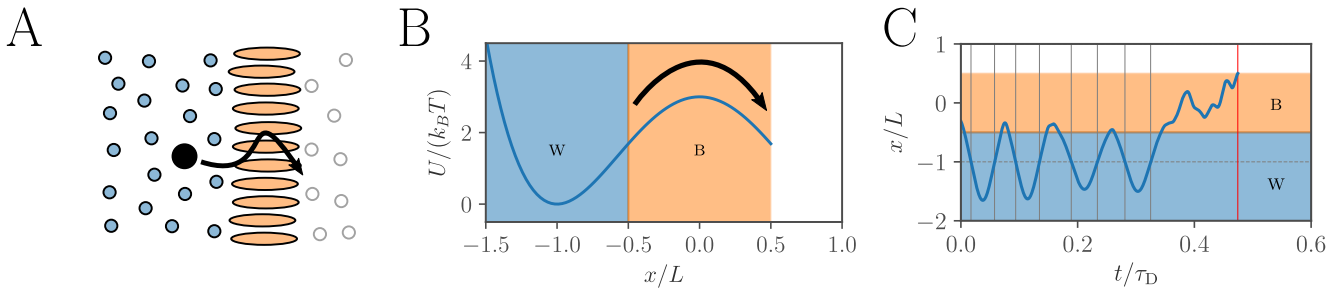

 \begin{overpic}[width=\textwidth]{{/../figs/posMem_intro3}.pdf}
 \put(0,20){\huge A}
 \put(29,20){\huge B}
 \put(65,20){\huge C}
 \end{overpic}
 \caption{{\bf A:} The dynamics of a particle diffusing through a membrane, separating different solvent species serves as an illustrative example for space-inhomogeneous friction effects. {\bf B:} The truncated quartic potential eq.~\eqref{eq:double_well_potential} is used to model barrier-crossing dynamics. The colors indicate regions $X_i$ with different local friction. {\bf C:} Example trajectory, simulated using single-exponentially decaying locally-coupled memory-friction components, eq.~\eqref{eq:exp_kernel} with $\tau_{\mathrm{W}}/\td=1$ and $\tB/\td=0.01$, that together form the friction kernel of eq.~\eqref{eq:kernel_local}. The coupling function $\chi_i(x)$ is given by eq.~\eqref{eq:chi} for the regions $X_i$, also shown in B. The \acl{MFPT} (\acs{MFPT}) for barrier crossing is defined as the average of all time differences between crossings of the well bottom at $x/L=-1$ (shown as gray vertical lines) to the escape at the boundary at $x/L=0.5$ (shown as a red vertical line).}
 \label{barrier_intro}
\end{figure*}

If the dissipative coupling between reaction coordinate and heat bath is assumed linear, the dynamics is described by an approximate version of the \ac{GLE}, with memory effects due to the finite relaxation time of the heat bath \cite{Zwanzig1960,Mori1965a}.
However, in many physical systems the dissipative interaction between reaction coordinate and heat bath  depends non-linearly on the current state of the reaction coordinate \cite{Zwanzig1973}.
For example, a small molecule traversing a membrane separating two different fluids, as
illustrated in fig.~\ref{barrier_intro}A, 
clearly interacts with different orthogonal degrees of freedom, namely fluid or membrane molecules, 
depending on 
where it is currently located.
As a second example, a reaction coordinate describing the folding of a protein is expected to experience different friction depending on whether the protein is unfolded or folded.
Even for a single confined solute particle in a fluid, the non-linear dissipative interaction of the particle and its surrounding fluid molecules leads to confinement-dependent memory effects \cite{Daldrop2017a}; for colloidal particles in a viscoelastic fluid, such non-linear solute-solvent interactions have been observed experimentally \cite{Muller2020}.

The first analytical relation between the friction magnitude and the barrier-crossing
 time was derived by
  Kramers \cite{Kramers1940}.
Kramers considered the memory-less, i.e. Markovian, Langevin equation with homogeneous friction magnitude. He showed that,
while in the high-friction limit the \ac{MFPT} scales linearly with the friction,
in the low-friction scenario the \ac{MFPT} scales linear with the inverse of the friction magnitude.
The crossover between these two asymptotic results was eventually bridged by a theory due to
Melnikov and Meshkov \cite{Melnikov1986} (MM), which is valid for all values of the friction magnitude.

For the scenario where there is no time-scale separation between heat bath and reaction coordinate, so that non-Markovian memory effects are relevant, the first theory to describe barrier-crossing times is due to Grote and Hynes (GH) \cite{Grote1980}. 
 In their theory, only the local memory effects in the barrier region are taken into account, and away from the barrier region the reaction coordinate is assumed Markovian. 
For the case of homogenous memory effects throughout the well and barrier regions, Pollack, Grabert, and H\"anggi (PGH) derived the \ac{MFPT}, which for long memory, scales quadratically with the memory time, so that the former can exceed the latter by orders of magnitude \cite{Pollak1989}. 
For the special case of a single-exponential memory function, the quadratic scaling of the \ac{MFPT} with memory time was also derived analytically from a harmonic approximation, and a simple heuristic formula which reproduces the results of PGH theory was proposed \cite{Kappler2018}.
Importantly, for systems with the same long memory, the predictions of the barrier-crossing time by PGH and GH theories can differ by many orders of magnitude. Evidently it is crucial whether the coupling between reaction coordinate and heat bath is linear throughout well and barrier region (homogeneous friction) or non-linear and thus different in well and barrier region (space-inhomogeneous friction)  \cite{Talkner1988,Kappler2018}.

For homogeneous friction, there exist numerical studies of barrier crossing considering  both single-timescale memory  \cite{Straub1986,Tucker1991,Ianconescu2015,Daldrop2018,Kappler2018}, 
as well as the implications of several memory time scales \cite{Kappler2019,Lavacchi2020}.
Models incorporating space-inhomogeneous friction have so far mostly been studied in the double limit where inertial and memory effects are negligible and can thus be modeled via an overdamped Langevin equation \cite{Berezhkovskii2011},
or the equivalent Fokker-Planck 
equation \cite{Hummer2004,Hinczewski2010}. 
However, this limit is subtle, as non-Markovian memory effects can generate spurious space-inhomogeneous friction if interpreted in terms of a Markovian model \cite{Ayaz2021}.
For space-inhomogeneous memory friction magnitude with a single homogeneous timescale, some works observed significant deviations of the \ac{MFPT} in both analytic theory and simulations \cite{Straus1993, Pollak1993, Haynes1994}.
The implications of space-inhomogeneous memory time scales in the well and barrier regions on the global barrier-crossing dynamics, have so far only been addressed by an analytical model  \cite{Singh1990, Krishnan1992, Singh1994}
which is able to theoretically bridge the GH and PGH scenarios. However, this model has never been challenged by numerical simulations.

We here present a model system to study space-inhomogeneous friction memory times and magnitudes, which in the appropriate parameter regimes reproduces the predictions of both PGH and GH theory.
Our model is based on the non-linear Zwanzig model \cite{Zwanzig1973,Krishnan1992}. Importantly, while the model makes certain simplifying assumptions that are not guaranteed to hold for general systems, it allows us to study under which conditions the \ac{MFPT} is determined dominantly by either the memory friction in the well or in the barrier region. 
Specifically, we consider a reaction coordinate subject to a potential well, bounded by a moderate barrier on one side, as illustrated in fig.~\ref{barrier_intro}B. 
In the well and barrier regions the reaction coordinate is locally coupled to independent heat baths, each with a single finite and in general different relaxation time.
This local coupling leads to space-inhomogeneous single-exponential memory in the reaction coordinate, and by independently varying the memory effects in well and barrier region we disentangle the effects of space-inhomogeneous memory times, $\tW$ for the well and $\tB$ for the barrier, and friction magnitudes, $\gW$ and $\gB$, on the barrier-crossing time.
By comparing results of numerical simulations to the rate theories 
of GH theory \cite{Grote1980} and PGH theory \cite{Pollak1989} (for which we for simplicity use the heuristic formula \cite{Kappler2018}), 
with the latter evaluated using either the well or barrier friction, we are able to infer which theory describes the numerical results, and whether the barrier-crossing time depends dominantly on the well or barrier friction.

We present the results of our numerical study in two parts. 
First we discuss the Markovian regime, for which memory effects in both the well and barrier regions are negligibly small, i.e. $\tB$ and $\tW$ are much smaller than the diffusive timescale $\td$. 
The dynamics in this regime are thus dependent only on inertial effects, which, strictly speaking, are only Markovian if both instantaneous position and velocity are used for defining a configuration. By labeling inertial effects as Markovian, we demarcate such inertial effects from non-Markovian effects due to coupling of the principle coordinate with hidden heat bath degrees of freedom.
In our model inertial effects are characterized by the inertial time scale for the different friction magnitudes, $\tau_{m,\mathrm{B}}=m/\gB$ or $\tau_{m,\mathrm{W}}=m/\gW$. 
We find that whenever the well dynamics is in the high-friction regime, $m/\gW \ll \td$, then the barrier-crossing time is determined by the barrier top friction. 
If then, the barrier top is also in the high-friction regime, $m/\gB \ll \td$, the \ac{MFPT} is described by Kramers theory \cite{Kramers1940}, evaluated using the friction magnitude at the barrier top. 
If instead, the barrier dynamics is in the low-friction regime, $m/\gB > \td$, while $\gB$ and $\gW$ are not too different, then MM theory \cite{Melnikov1986} or PGH theory \cite{Pollak1989}, evaluated using the barrier parameters, describe the numerically obtained barrier-crossing times.
On the other hand, if the dynamics in the well is in the low-friction regime, $m/\gW > \td$, then the \ac{MFPT} is described by PGH theory or MM theory, both evaluated using the well parameters, which therefore dominates the global barrier-crossing dynamics.

In the second part we discuss the non-Markovian regime, where memory effects in either, well or barrier regions, are relevant, i.e.~either $\tB$ or $\tW$ are of at least similiar order as the diffusive timescale $\td$.
For simplicity we keep the friction magnitudes equal, $\gB = \gW$, and in the high-friction regime by imposing $m/\gB = m/\gW \ll \td$.
Analogous to the previous case, we find that whenever the well dynamics is in the Markovian regime, $\tW \ll \td$, then the barrier-crossing time is determined by the barrier top friction.  
If then again, the barrier top is also in the Markovian regime, $\tB \ll \td$, the \ac{MFPT} is described by Kramers theory \cite{Kramers1940}, evaluated using the friction parameters for the barrier top. 
If instead, memory effects in the barrier region are relevant, then GH theory \cite{Grote1980} agrees with the numerically obtained barrier-crossing times.
In contrast, if the well memory is long, $\tW >\td$, the \ac{MFPT} is described by PGH theory using the well parameters.
While then in general the \ac{MFPT} is rather independent of the barrier friction, Markovian barrier dynamics lead to a speed up of \acp{MFPT} as compared to a barrier region with long memory.
This speedup, which we term 'Markovian-barrier acceleration',  is not captured by any presently available rate theory, but can be understood from a committor analysis, analogous to the `non-Markovian-barrier acceleration' already predicted by GH theory.

The remainder of this paper is organized as follows.
In section II we first introduce the space-inhomogeneous memory model we consider.
In section III, we then compare numerical simulations of our model 
to rate-theory predictions. We first consider the short-memory limit,
and subsequently study how local memory effects modify the \ac{MFPT}.
In our concluding section IV, we provide a table which 
summarizes our results.

\section{Model}

We consider a reaction coordinate $x$ and $N$ non-interacting heat baths with finite relaxation dynamics  \cite{Zwanzig1973, Krishnan1992, Haynes1994};
for $x \in X_i$, the reaction coordinate couples linearly to the $i$-th heat bath.
As we show in appendix \ref{sectionDerivations}\ref{subs_gleDeriv},
 integrating out the bath degrees of freedom then leads to 
a \ac{GLE}
\begin{align}
\label{eq:gle_local}
\begin{split}
m\ddot{x}(t)=&-\int_0^{t} dt'\, \Gamma[t-t',x(t),x(t')]\dot{x}(t')\\
&-\ddx U[x(t)]+\noiseLoc,
\end{split}
\end{align}
which is a generalization of the model proposed by \citet{Zwanzig1973}. $U[x(t)]$ is a potential landscape, and the space- and time-dependent friction kernel $\Gamma\left[t-t',x(t),x(t')\right]$ is given as 
\begin{align}
\label{eq:kernel_local}
\Gamma\left[t-t',x(t'),x(t)\right]= \sum_{i=1}^N \chi_i[x(t)] \Gamma_i(t-t') \chi_i[x(t')],
\end{align}
where the purely time-dependent components $\Gamma_i$ describe the internal relaxation
dynamics of reservoir $i$, and the dimensionless functions $\chi_i$, defined by
\begin{align}
\label{eq:chi}
\chi_i(x) = \begin{cases}1 &\mathrm{if}~x\in X_i, \\ 0 &\mathrm{if}~x \notin X_i,\end{cases}
\end{align}
describe the coupling of the reaction coordinate $x$ to reservoir $i$.
The terms in eq.~\eqref{eq:kernel_local} have a simple intuitive interpretation:
at any past time $t'$, the reaction coordinate $x$
perturbs reservoir $i$ via the coupling strength $\chi_i[x(t')]$;
this perturbation relaxes in the heat bath for a duration $t-t'$ as described by $\Gamma_i(t-t')$, and
finally couples back to the reaction coordinate at the time $t$ via $\chi_i[x(t)]$.

As we show in appendix \ref{sectionDerivations}\ref{subs_fdt}, the random force fulfills the fluctuation-dissipation relation
\begin{equation}
\label{eq:FDT_main}
\beta \langle \noiseLoc \eta[x(t'),t'] \rangle=\Gamma[t-t',x(t),x(t')],
 \end{equation}
 where $\beta^{-1} = k_{\mathrm{B}}T$ is the thermal energy with $k_{\mathrm{B}}$ the Boltzmann
constant and $T$ the absolute temperature.

For our numerical simulations we consider barrier crossing in the quartic potential 
\begin{equation}
\label{eq:double_well_potential}
U(x)= U_0\left[\left(\frac{x}{L}\right)^2-1\right]^2,
\end{equation}
with a length scale $L$ and barrier height $\beta U_0=3$ (we show some results with varying barrier heights in appendix \ref{sectionFurtherNumData}\ref{sectionBarrierHeight}).
To systematically study the effect of space-inhomogeneous memory on the \ac{MFPT}, 
we consider $N=2$ independent heat baths with coupling regions
in the well,
$X_{\mathrm{W}}/L=(-\infty,-0.5)$,
and on the barrier,
$X_{\mathrm{B}}/L=[-0.5,0.5)$,
as illustrated in fig.~\ref{barrier_intro}B.
For the resulting two memory kernels $\GW$ and $\GB$, we 
consider single-exponential kernels, 
\begin{equation}
\label{eq:exp_kernel}
	\Gamma_i(t)=\frac{\gamma_i}{\tau_i}e^{-t/\tau_i},
\end{equation}
 with friction magnitudes $\gamma_i$ and 
 relaxation time scales $\tau_i$, where $i \in \{\mathrm{W},\mathrm{B}\}$.
This means that the particle interacts with two independent heat baths, each
of which relaxes according to a single exponential.

As we show
 in appendix \ref{sectionDerivations}\ref{subs_gleDL}, eq.~\eqref{eq:gle_local} with local memory eq.~\eqref{eq:kernel_local}
 can be cast into dimensionless form 
by introducing a diffusion time scale $\td=\beta \gamma L^2$ with $\gamma=\sum_i \gamma_i$, 
and an inertial time scale $\tm=m/\gamma$.
With the potential eq.~\eqref{eq:double_well_potential} and a given barrier height $\beta U_0$, 
the system is then specified by four dimensionless parameters which we choose to be
the dimensionless inertial time scale $\tau_m/\tau_D$, 
the dimensionless local memory times $\tau_i/\tau_D$, $i \in \{W, B\}$,
and one of the two relative friction magnitudes $\gamma_i/\gamma$, $i \in \{W, B\}$.
To transform dimensionless results to physical dimensions, the temperature $T$,
the length scale $L$, and the sum $\gamma$ of the local friction magnitudes additionally need to be specified.
To simulate the dimensionless formulation of eqs.~\eqref{eq:gle_local}, \eqref{eq:kernel_local}, \eqref{eq:FDT_main}, we use a {Markovian embedding} whereby we explicitly simulate the dynamics in the reservoirs, as detailed in appendix \ref{sectionDerivations}\ref{subs_gleDL}.

In analogy to previous works, we define various limits by comparison of respective time scales with the diffusive time scale $\td$. For example the Markovian limit where memory effects are negligible is obtained for $\tau_i < \td$, and the high-friction limit where inertial effects are negligible is obtained for $\tm < \td$ \cite{Kappler2018, Kappler2019, Lavacchi2020}.
However, since the local friction in region $i$, $\gamma_i$, is only part of the total friction $\gamma$, which is used to define $\td$, a condition $\tau_i \ll \td$ does not automatically ensure the expected limit in region $i$.
Rather, a condition involving the local diffusive time scale, $\tau_{D,i} \equiv \beta L^2 \gamma_i$, needs to be used, namely $\tau_i/\tau_{D,i} \equiv \tau_i/\tau_D \cdot \gamma/\gamma_i \ll 1$.
Similarly,  inertial effects are locally relevant for $\tmi/\tdi \equiv \tm/\td \cdot (\gamma/\gamma_i)^2 \gtrsim 1$, where
$\tmi \equiv m/\gamma_i$.
While the distinction between $\tmi$, $\tdi$ and $\tm$, $\td$ is important
if $\gamma_i$ is significantly smaller than $\gamma$, 
for most of the parameter combinations  
we consider in the main text, all $\gamma_i$ 
are of similar order as their sum $\gamma$.

{Since we are interested in evaluating the barrier-crossing 
time starting from the well region $X_{\mathrm{W}}$, 
we restart the simulation once the particle crosses the right boundary at $x/L =0.5$, which we consider 
a successful escape. 
At each restart, we draw the initial position from an approximate 
 Boltzmann distribution around the well minimum, 
i.e.~we draw a Gaussian random variable $x(0)$ with $\langle x(0)/L \rangle = -1$ 
and  $\langle (x(0)/L + 1)^2 \rangle = k_BT/(L^2 U''(-L))= k_BT/(8U_0)$.
Similarly, we draw the initial velocity $\dot x(0)$ from its equilibrium distribution, i.e.~a Gaussian random
variable with $\langle \dot x(0) \rangle = 0$ and  $\langle \dot x^2(0) \rangle = k_BT/m$.
The initial conditions for the heat bath variables we subsequently draw from their respective Boltzmann distributions, as detailed in appendix \ref{sectionDerivations}\ref{subs_fdt}.}

In fig.~\ref{barrier_intro}C we show an example trajectory, simulated using long memory in the well and short memory on the barrier, for which different dynamics in the different regions are clearly observed:
While in the well region the trajectory oscillates weakly damped around the potential minimum, 
on the barrier the trajectory is more akin to overdamped diffusive dynamics.
Figure \ref{barrier_intro}C furthermore illustrates how we compute the \ac{MFPT} 
from observed time differences between crossings of the well minimum and the escape at $x/L=0.5$.
That this is a reliable method for calculating the \ac{MFPT} has been shown before \cite{Kappler2018}.
 
In the main text, we compare our numerical results to GH \cite{Grote1980}, MM \cite{Melnikov1986}, and PGH theory \cite{Pollak1989}, where instead of the latter we use the heuristic formula \cite{Kappler2018} in practice.
GH theory only accounts for barrier memory friction, which is why we always evaluate it using the barrier memory kernel; the theory assumes fast equilibration within the well, and does not depend on the well friction explicitly.
Both the Markovian MM and non-Markovian PGH theory assume homogeneous friction. 
We therefore evaluate these theories using either the local parameters $\gamma_i$, $\tau_i$ of the well or barrier region.
This allows us to infer not only which rate theory describes the barrier-crossing dynamics in which regime,
but also which region (well/barrier) dominantly determines the global \acp{MFPT}. 
In appendix \ref{rateTheorySection} we summarize the equations used to calculate predictions
for all rate theories considered in the main text.
In the main text, we do not compare our numerical simulations to the analytical rate theory for space-inhomogeneous memory friction due to \citet{Krishnan1992}.
The reason for this is twofold: First, by comparing to the widely used GH and PGH theories, we are able to assess which local dynamics dominate the global \ac{MFPT}. 
Second, as we show in appendix \ref{sectionFurtherNumData}\ref{sectionKSR}, the theoretical predictions by \citet{Krishnan1992} do not capture the `Markovian-barrier acceleration' regime which we prominently discuss below, and which we quantify using the PGH predictions.

\section{Results}

\begin{figure*}
\centering
 \begin{overpic}[width=0.4\textwidth]{{/../figs/barrierReset_flo_pub/TBLR0.000_overTM_varGB_MM_MM}.pdf}
 \put(-2,62){\huge A}
 \end{overpic}
\begin{overpic}[width=0.4\textwidth]{{/../figs/barrierReset_flo_pub/TBLR0.000_overGB_varTM_MM}.pdf}
 \put(2,62){\huge D} 
 \end{overpic} 
 \begin{overpic}[width=0.4\textwidth]{{/../figs/barrierReset_flo_pub/selTraj_GB0.9_TM0.000_TLR0.000_TB0.000}.pdf}
 \put(-2,62){\huge B}
 \end{overpic}
 \begin{overpic}[width=0.4\textwidth]{{/../figs/barrierReset_flo_pub/map_TLR0.000_TMoverGB}.png}
 \put(2,62){\huge E} 
 \end{overpic}
 \begin{overpic}[width=0.4\textwidth]{{/../figs/barrierReset_flo_pub/selTraj_GB0.9_TM10.000_TLR0.000_TB0.000}.pdf}
\put(-2,62){\huge C}
 \end{overpic}
\begin{overpic}[width=0.4\textwidth]{{/../figs/mfptOD_barrier_model}.pdf}
\put(2,62){\huge F}
\put(82,58){\large F1}
\put(82,35){\large F2}
\end{overpic} 
 \caption{\Acl{MFPT} (\acs{MFPT}), $\tmfp/\td$, for different barrier memory friction, $\GB(t)=\gB/\tB e^{-t/\tB}$ and well memory friction, $\GW(t)=\gW/\tW e^{-t/\tW}$, compared with analytical predictions given by \citet{Melnikov1986} (MM). The data is shown for various inertial time scales $\tm/\td$ and equal memory times in the Markovian limit, $\tB/\td=\tW/\td=10^{-4}$. \textbf{A:} \Ac{MFPT} plotted over the inertial time scale $\tm/\td$ for different ratios of the barrier to total friction magnitude $\tB/\gamma$. The predictions by \ac{MM} are shown for the effective barrier-friction paramters given by $\gB$ and for the effective well-friction parameters given by $\gW$. {\bf B, C:} Example trajectories. \textbf{D:} \Ac{MFPT} plotted over $\tB/\gamma$ for various $\tm/\td$. 
 The predictions by \ac{MM} are shown for the effective barrier-friction paramters as broken lines and for the effective well-friction parameters as solid lines. \textbf{E:} Color map of agreement of the simulation results with the theoretical predictions. The color denotes whether the simulated $\tmfp \in [0.5\,\tau_{\mathrm{theo.}} , 2\,\tau_{\mathrm{theo.}} ]$, where $\tau_{\mathrm{theo.}}$ is calculated using \ac{MM} theory and either the effective well- or barrier-friction parameters. The hatching indicates that both theoretical predictions agree with the simulated data. 
{\bf F:} Model potential (F1) and friction profile (F2) considered to study the effect of the barrier friction on the \ac{MFPT} in the high-friction Markovian regime. 
For this model the \ac{MFPT} is exactly given by eq.~\eqref{eq:mfptOD_barrier}.
 }
\label{barrierResults_G}
\end{figure*}

 In order to decouple Markovian inertial effects
 and non-Markovian memory effects
we analyze both scenarios independently. For this we first
consider the Markovian limit and vary $\tm/\td$,
and second the high-friction limit, $\tm/\td \ll 1$, with varying $\tB/\td$ and $\tW/\td$.

\subsection*{Markovian friction dynamics}

We now consider the Markovian limit for both well and barrier.
In fig.~\ref{barrierResults_G}A  we
show the rescaled \ac{MFPT} $\tmfp/\td$ as a function of
the rescaled inertial time
$\tm/\td$.
For reference, we  include numerical results from a GLE with a homogeneous single-exponential memory kernel
with memory time $\tau_{\mathrm{glob.}}/\td = 10^{-4}$ and a single friction magnitude $\gamma_{\mathrm{glob.}} = 0.9\, \gamma$ (chosen as to coincide with the blue solid line, as explained further below) \cite{Kappler2018};
the resulting MFPTs are shown in fig.~\ref{barrierResults_G}A as gray triangles,
and clearly show the Kramers turnover between high-friction dynamics for $\tm/\td \ll 1$, 
 where the \ac{MFPT} scales as $\tmfp \sim \gamma$,
 and  low-friction dynamics for $\tm/\td \gg 1$, where the \ac{MFPT} scales 
 as $\tmfp \sim m/\gamma$  \cite{Kramers1940, Melnikov1986, Kappler2018}.

Figure \ref{barrierResults_G}A furthermore
shows numerical results for the space-inhomogeneous memory model 
eqs.~\eqref{eq:gle_local} and \eqref{eq:kernel_local}
for $\tB/\td=\tW/\td = 10^{-4}$ and the two values $\gB/\gamma= 0.1, 0.9$.
For $\gB/\gamma = 0.1$ we have $\gW/\gamma = 0.9$, so that the friction in the well is almost one order of magnitude larger as compared to the friction in the barrier region.
Conversely, for $\gB/\gamma = 0.9$ the friction in the well, $\gW/\gamma = 0.1$, is almost one order of magnitude smaller as compared to the friction in the barrier region.
While in the high-friction regime $\tm/\td \ll 1$, 
the results for $\gB/\gamma= 0.9$ (blue circles; barrier friction much larger than well friction) agree well with the global memory friction data (gray triangles),
in the low-friction regime $\tm/\td \gg 1$, it is the \ac{MFPT} for $\gB/\gamma = 0.1$ 
(orange squares; well friction much larger than barrier friction) 
that is comparable to the global memory friction result.
This indicates that for high friction, the \ac{MFPT} is dominated by the barrier friction, 
whereas for low friction the \ac{MFPT} is dominated by the well friction.

The crossover between barrier-dominated \ac{MFPT} to well-dominated \ac{MFPT}
observed in fig.~\ref{barrierResults_G}A 
is further confirmed by comparing the numerical data to predictions of the \ac{MM} theory for Markovian barrier crossing, which is based on homogeneous friction.
In fig.~\ref{barrierResults_G}A, we show the predictions of \ac{MM} theory, evaluated using either the well friction $\gW$ or the barrier friction $\gB$.
Note that because of the symmetry in the used parameters, the blue solid line represents both the \ac{MM} prediction for $\gB/\gamma = 0.9$, and evaluation using the barrier friction, as well as the \ac{MM} prediction for $\gB/\gamma = 0.1$, and evaluation using the well friction.
On the other hand, the orange broken line represents the opposite parameter choice in both scenarios.
We observe that, while in the high-friction limit $\tm/\td \ll 1$, 
the simulated \acp{MFPT} agree with the \ac{MM} predictions evaluated 
at the barrier region, for low friction $\tm/\td \gg 1$ the 
numerical data is described by the \ac{MM} theory evaluated
at the well.

That for the Markovian high-friction scenario,
 the \ac{MFPT}  is dominated by the barrier friction, is also appreciated by a simple analytical model.
In the high-friction Markovian limit the \acl{MFPT} $\tmfp$ to start at $x_0$ and reach $x_f$ in a potential $U(x)$ and for space-inhomogeneous friction $\gamma(x)$  is derived exactly from the Fokker-Planck equation 
as \cite{Hinczewski2010}
\begin{align}
\label{eq:mfptOD}
\tmfp(x_0) = \beta \int_{x_0}^{x_f} dx\ \gamma(x) e^{-\beta U(x)} \int_{x_L}^{x} dx'\ e^{-\beta U(x')},
\end{align}
where $x_L< x_0$ is a lower reflecting boundary. 
To study the effect of barrier friction on the \acl{MFPT}, we consider the model illustrated in fig.~\ref{barrierResults_G}F: a simplified flat potential $U(x)$, which features a reflecting boundary at $x=0$ and a step barrier of height $U_0$, width $B$ and friction $\gB$ located at position $x=L/2$. Outside of the potential barrier, the friction is $\gW$. 
Considering $x_0=x_L=0$  and $x_f=L$, the \acl{MFPT} is calculated from eq.~\eqref{eq:mfptOD}  to be
\begin{align}
\label{eq:mfptOD_barrier}
\tmfp^{0 \to L} &= \beta \frac{L^2}{2} \gW  + \beta \frac{LB}{2} \left( \gB - \gW \right) \\
&\quad + \beta \frac{B(L-B)}{2} \left( 1- e^{-\beta U_0}\right) \left(\gB e^{\beta U_0} - \gW \right).
\nonumber
\end{align}
In the high-barrier limit, where $\beta U_0 \gg 1$, eq.~\eqref{eq:mfptOD_barrier} is dominated by an expression which only contains the barrier friction $\gB$
\begin{align}
\label{eq:mfptOD_barrier_limit}
\tmfp^{0 \to L} \approx  \beta \gamma_B \frac{B(L-B)e^{\beta U_0}}{2}.
\end{align}
This result explains why the MFPT in the high-friction scenario is determined by the barrier friction.

 Example trajectories, comparing the cases of Markovian high-friction dynamics ($\tm/\td=10^{-4}$), where the MFPT is determined by the barrier friction, 
and Markovian low-friction dynamics ($\tm/\td=10$), where
the MFPT is determined by the well-friction, are shown in figs.~\ref{barrierResults_G}B and C. 
While the trajectory in fig.~\ref{barrierResults_G}B generally exhibits dynamics reminiscent of
Markovian high-friction Langevin dyamics around the well and also in the barrier region, 
differences in the lengths of persistent motion due to the vastly different local friction magnitudes are clearly visible. 
The trajectory in fig.~\ref{barrierResults_G}C shows oscillations within the wells and long residence times, 
which are typical of inertia-dominated stochastic dynamics \cite{Kappler2018}.

Figure~\ref{barrierResults_G}D shows numerical MFPTs, plotted as a function of 
the relative barrier friction $\gB/\gamma$ for various values of the rescaled inertial time $\tm/\td$. 
Again, while for high friction, $\tm/\td=10^{-4}$,
 the simulated data agrees with the \ac{MM} theory evaluated using the barrier friction (orange broken line), 
 for large $\tm/\td=10$ the numerical results agree with the predictions using the 
  well friction (blue solid line). 
We observe that for the parameters considered, the rescaled \ac{MFPT} $\tmfp/\td$ always increases monotonously with $\gB/\gamma$
indicating that increasing barrier friction while decreasing well friction slows down barrier-crossing.
The analytical \ac{MM} theory shows non-monotonicities for the case of high total friction, $\tm/\td=10^{-4}$, but very unequal friction magnitudes in well and barrier regions, $\gamma_{i}/\gamma \ll 1$, i.e. to the far right and left of fig.~\ref{barrierResults_G}D. 
This is discussed in detail in appendix \ref{sectionFurtherNumData}\ref{sectionInertialBarrier}.

\begin{figure*}[htb]
  \begin{overpic}[width=0.4\textwidth]{{/../figs/barrierReset_flo_pub/TM0.000_overTB_varTLRcompSingle}.pdf}
\put(-2,62){\huge A}
 \end{overpic}
  \begin{overpic}[width=0.4\textwidth]{{/../figs/barrierReset_flo_pub/TM0.000_overTLR_varTB}.pdf}
 \put(2,62){\huge D}
 \end{overpic}
  \begin{overpic}[width=0.4\textwidth]{{/../figs/barrierReset_flo_pub/selTraj_GB0.5_TM0.000_TLR0.000_TB1.000}.pdf}
\put(-2,55){\huge B}
 \end{overpic} 
  \begin{overpic}[width=0.4\textwidth]{{/../figs/barrierReset_flo_pub/map_TM0.000_TLRoverTB}.png}
 \put(2,62){\huge E}
 \end{overpic} 
   \begin{overpic}[width=0.4\textwidth]{{/../figs/barrierReset_flo_pub/selTraj_GB0.5_TM0.000_TLR1.000_TB1.000}.pdf}
\put(-2,55){\huge C}
 \end{overpic} 
  \begin{overpic}[width=0.4\textwidth]{{/../figs/barrierReset_flo_pub/TPInVel_TM0.000_TLR1.000_varTB}.pdf}  
\put(-2,62){\huge F}
 \end{overpic}
 \caption{\Acl{MFPT} (\acs{MFPT}), $\tmfp/\td$, for different barrier memory friction, $\GB(t)=\gB/\tB e^{-t/\tB}$ and well memory friction, $\GW(t)=\gW/\tW e^{-t/\tW}$, compared with analytical predictions given by \citet{Grote1980} (GH, broken lines) and \citet{Pollak1989} (PGH, solid lines, evaluated using the heuristic formula \cite{Kappler2019}). The data is shown for various barrier-friction $\tB/\td$ and well-friction times $\tW/\td$, constant inertial time scale in the high-friction limit $\tm/\td=10^{-4}$ and equal friction magnitudes $\gB/\gamma = \gW/\gamma = 0.5$. {\bf A:} \Ac{MFPT} plotted over the barrier-friction time $\tB/\td$. The theories are shown for the respective barrier-friction time in gray and in case of PGH theory for the well-friction time as colored lines. \Acp{MFPT} to reach the barrier entry at $x/L=-1/2$ are shown as colored dash-dotted lines. {\bf B, C:} Example trajectories for the barrier-dominated and GH-predicted limit (B) and the well-dominated PGH-predicted limit (C). {\bf D:} \Ac{MFPT} plotted over the well-friction time $\tW/\td$. The theories are shown for the respective well-friction time in gray and in case of the GH for the barrier-friction time as colored broken lines. \textbf{E:} Contour plot of agreement of the simulation results with the theoretical predictions. The color denotes whether the simulated $\tmfp \in [1/3\,\tau_{\mathrm{theo.}} , 3\,\tau_{\mathrm{theo.}} ]$, where $\tau_{\mathrm{theo.}}$ is calculated using either the GH theory with the barrier-friction parameters or the PGH theory with the well-friction parameters. The hatching indicates that both theoretical predictions agree with the simulated data. The light blue area denotes the `Markovian-barrier acceleration' of the PGH prediction for which we define $\tau_{\mathrm{theo.,MBA}}=0.2\,\tau_{\mathrm{theo.,PGH}}$. {\bf F:} Committor $p(\text{TP}|v_{\rm in})$ for transition paths crossing the barrier region plotted over the initial velocity $v_{\rm in}$ upon entering the barrier region for various barrier-friction times $\tB/\td$ and constant well-friction times $\tW/\td = 1$ and inertial times $\tm/\td = 10^{-4}$. The velocity related to the difference in potential energy relative to the barrier top is plotted as a vertical black broken line. The flux-weighted equilibrium velocity distribution $p_{\rm eq}(v_{\mathrm{in}})$ is plotted as a gray broken line on a linear scale.
 }
 \label{barrierResults_T}
\end{figure*}
 
Figure~\ref{barrierResults_G}E illustrates for which parameters the simulated \ac{MFPT} is described by  the theoretical predictions of \ac{MM} theory, evaluated for either the well- or barrier-friction parameters.
The figure again clearly shows that for high-friction dynamics, $\tm/\td \ll 0.1$, the \ac{MFPT} is determined
by the barrier friction, whereas for low-friction dynamics, the well friction determines
the \ac{MFPT}.
The hatched area shows the overlap where both predictions calculated using well or barrier friction agree with the simulated \ac{MFPT}. 
Obviously, in the crossover between barrier- and well-dominated friction, where $\gB \approx \gW$,
 the rate theories produce similar results when evaluated using barrier- or well-friction,
  see also fig.~\ref{barrierResults_G}A.
This is because for $\gB/\gamma \approx 0.5$, we have $\gW/\gamma = (\gamma-\gB)/\gamma \approx 0.5$,
so that the effective friction magnitudes in well- and barrier region, and hence the predictions
of \ac{MM} theory, which depend on the effective local friction, are similar.

To summarize figs.~\ref{barrierResults_G}A--E,
in the Markovian (short memory) limit, the rescaled \ac{MFPT}
is for high-friction dynamics determined by the barrier friction, whereas for low-friction dynamics it is determined by the well friction.
The former effect is illustrated by the analytical result eq.~\eqref{eq:mfptOD_barrier_limit}, while the latter is intuitively understood from the concept of energy diffusion.
For low-friction dynamics the energy exchange between the reaction coordinate and the heat bath is weak and therefore the energy to cross the potential barrier is only slowly built up in the well region. 
This
process is  dominated by the well dynamics and
 leads to a slow-down of the global barrier-crossing times.
Since slow energy diffusion is also apparent for long memory times, a similar effect is observed in the discussion of the non-Markovian dynamics in the following.

\subsection*{Non-Markovian friction dynamics}

In fig.~\ref{barrierResults_T} we investigate the memory-time dependence of the \ac{MFPT}.
For this, we consider a constant  inertial time scale in the high-friction limit, $\tm/\td=10^{-4}$,
and identical friction magnitudes for the two reservoirs, $\gW/\gamma=\gB/\gamma=0.5$,
while varying the well- and barrier-friction time scales, $\tW/\td$ and $\tB/\td$.
We compare our numerical results to analytical predictions based on both \ac{GH} theory \cite{Grote1980}, 
which we evaluate using the barrier-friction parameters $\tB$, $\gB$
and which is hence independent of the well parameters,
and PGH theory \cite{Pollak1989} (for which we in practice use the heuristic formula \cite{Kappler2018}), 
 which we evaluate for both the well parameters $\tW$, $\gW$
or the barrier parameters $\tB$, $\gB$. 

In fig.~\ref{barrierResults_T}A we show the rescaled \ac{MFPT} as a function of
the barrier memory time $\tB/\td$ for various well memory times $\tW/\td$.
For short barrier memory, $\tB/\td \ll 0.1$, the dynamics on
the barrier top is Markovian and the 
 numerical \acp{MFPT} 
are independent of $\tB/\td$.
If additionally also the memory in the well is short, $\tW/\td \ll 0.1$, 
we are in the Markovian high-friction limit.
While, as we have already discussed in the context of fig.~\ref{barrierResults_G}A, 
in this limit the barrier-crossing time is 
determined by the barrier friction,
also PGH theory evaluated with well friction (gray solid line)
agrees with PGH theory evaluated with barrier friction (colored solid lines),
and GH theory (gray broken line; always evaluated at barrier friction);
this is because we have equal friction magnitudes in well and barrier.
While in the double limit of high friction and short well memory $\tW/\td \leq 0.01$,  
 all theories
 describe the numerical data as long as $\tB/\td \ll 0.01$, 
 for $\tB/\td \gtrsim 0.01$, shown as green and orange symbols, both GH theory and numerical results 
  display an acceleration (as 
  compared to the Markovian limit $\tB/\td \to 0$); we refer to this as GH 'non-Markovian-barrier acceleration'.
 That GH describes this acceleration regime is expected because
GH theory was derived assuming fast equilibration within each well, 
which is in line with
 the high-friction Markovian dynamics inside the well for $\tm/\td = 10^{-4}$, $\tW/\td \leq 0.01$. We note that in the limit of $\tB/\td \rightarrow \infty$, GH theory agrees with the predictions of transition state theory \citep{Grote1980}.
For high-friction dynamics with short well memory,
 the \ac{MFPT} is thus determined by the barrier friction and described by GH theory.

On the other hand, for long memory in the well, $\tW/\td \geq 1$, the numerical \acp{MFPT} 
(red diamonds and blue triangles)
are comparable to the predictions of PGH theory evaluated 
on the well parameters, and even for a Markovian barrier, $\tB/\td \ll 1$, disagree considerably with GH theory 
(which indicates a breakdown of the assumption of fast equilibration within the well which is assumed in GH theory).
In contrast to the Markovian-well scenario, for long well memory the \ac{MFPT}
is thus determined by the well dynamics.
Interestingly, in the long-well-memory regime, $\tW/\td \geq 1$,
the observed \acp{MFPT} show a slight acceleration for small barrier-memory 
times $\tB/\td \lesssim 0.1$, as compared to the predictions by PGH theory and the numerical results 
for $\tB/\td \gg 0.1$.
We here refer to this acceleration due to short barrier memory as
 `Markovian-barrier acceleration', which notably behaves opposite as a function
 of $\tB$ as compared to \ac{GH} `non-Markovian-barrier acceleration',
 as we discuss further below.

Figs.~\ref{barrierResults_T}B and C show example trajectories
where the \ac{MFPT} is determined by barrier or well friction.
For the trajectory shown in fig.~\ref{barrierResults_T}B the \ac{MFPT}
is determined by the barrier friction, and we observe
 high-friction Markovian dynamics within the well and a direct transition path upon entering the barrier region.
The trajectory with well-friction-determined \ac{MFPT}, fig.~\ref{barrierResults_T}C,
on the other hand shows long residence times, 
and multiple attempts entering the barrier region before crossing over the barrier top, that are associated with energy-diffusion, i.e.~memory- or inertia-dominated
  trajectories \cite{Kappler2018}.

In fig.~\ref{barrierResults_T}D we show numerical \acp{MFPT} 
 as function of the well memory time
 $\tW/\td$, for several constant values of the barrier memory time $\tB/\td$.
 We again compare
  to theoretical predictions based on PGH and GH theory. 
  Similar to fig.~\ref{barrierResults_T}A we see that for short well memory, $\tW/\td \ll 0.1$,
  the \ac{MFPT} becomes independent of the well memory time so that the dynamics is governed by the barrier.
If additionally the barrier memory time is short, $\tB/\td \ll 0.1$, 
then the \ac{MFPT} is described by both PGH (evaluated in the well) and GH theory.
Increasing the barrier memory time $\tB/\td$ then leads to an acceleration of barrier crossing
as we discussed in fig.~\ref{barrierResults_T}A,
and as described by GH theory (leftmost datapoints in fig.~\ref{barrierResults_T}D).
For any value of the barrier memory $\tB/\td$, we observe that as the well memory is increased, for $\tW/\td \gtrsim 1$ an asymptotic long-memory regime 
with $\tmfp \sim \tW^2$ is reached \cite{Kappler2018}, which is well-described by PGH theory
 evaluated at the well.
Increasing the well memory time $\tW/\td$ thus has both a qualitatively and quantitatively very 
different effect as increasing the barrier memory time $\tB/\td$ where, as we observe
in fig.~\ref{barrierResults_T}A, the \ac{MFPT} slightly increases/decreases (depending on $\tW/\td$)
and then becomes independent of $\tB/\td$.

Figure~\ref{barrierResults_T}E summarizes the agreement of the simulated high-friction \acp{MFPT} 
with PGH theory, evaluated on well parameters, and GH theory which is always evaluated using 
the barrier parameters.
There we see that once the well memory becomes relevant, i.e.~for $\tW/\td \gtrsim 0.1$,
the \ac{MFPT} is approximately described by PGH theory evaluated at the well.
The 'Markovian-barrier acceleration' regime appears if memory in the well is relevant,
but in the barrier region the memory time is significantly shorter, i.e.~for $\tB/\td \ll \tW/\td$,
and is shown as light blue.
If well memory effects are negligible, i.e.~for $\tW/\td \ll 0.1$,
but  memory effects are relevant in the barrier region, $\tB/\td \gg 0.1$, then 
GH theory describes the numerical results.
If memory effects are negligible for both well and barrier region, $\tW/\td \ll 1$ and $\tB/\td \ll 0.1$,
then we are in the Markovian limit, where the barrier friction $\gB$
determines the \ac{MFPT}.
That in this regime both GH theory (evaluated at the barrier region) 
and PGH theory (evaluated at the well region) describe the numerical \ac{MFPT},
as indicated by the hatching,
can be rationalized by the fact that we use the same friction magnitude for well and barrier, $\gB = \gW$.

\subsection*{Markovian-barrier acceleration}

In order to gain intuition about the 'Markovian-barrier acceleration' regime, i.e.~the slight barrier-crossing speed-up
observed for $\tW/\td \gtrsim 1$, $\tB/\td \ll 1$ in fig.~\ref{barrierResults_T}A and D,
we perform a committor analysis, the results of which are shown in fig.~\ref{barrierResults_T}F. 
The committor shown in the figure is defined as the 
probability to be on a transition path through the barrier region, and plotted
as a function of  the initial velocity with which the particle enters the barrier region, $v_{\rm in}$. 
For comparison, the flux-weighted
equilibrium velocity probability density, $p_{\rm eq.}(v) \propto v \exp(-m v^2/(2 k_B T))$ \cite{Daldrop2016,Voter1985}, is given as a gray broken line. Furthermore,
we show as a vertical black broken line the
threshold velocity $\vthres$ with which an undamped particle entering the barrier region
crosses over the barrier top, 
so that $m\vthres^2/2=U_0-U(x=-L/2) = 7 U_0/16$, and hence $\vthres = \sqrt{7 U_0/(8 m)}$.

For short memory in the barrier region, $\tB/\td \ll 1$, the committor is relatively small and only very slightly increases with larger initial velocities (orange and green solid lines), indicating that the kinetic energy is quickly dissipated in the barrier region and the probability to perform a transition is approximately independent of 
the velocity with which the particle enters the barrier region.
For long barrier memory $\tB/\td \geq 1$, the committor remains almost zero for velocities  $v_{\rm in} \lesssim \vthres$, indicating that many trajectories that enter the barrier region will simply roll back into the well region. 
They initially do not have enough kinetic energy to cross the barrier top and the  energy exchange with the barrier heat bath is not fast enough to gain the missing energy. At $\vthres$, the committor  starts to  increase sharply and saturates at a value of 1, 
which means that virtually every trajectory that enters the barrier region with 
at least this kinetic energy performs a transition through the barrier region. 
This is consistent with a weak energy exchange of heat bath and reaction coordinate in the barrier region, 
where a trajectory traverses the barrier top only if initially it has enough kinetic energy to reach there.

The physical picture for the 'Markovian-barrier acceleration' regime is hence that for short barrier memory time the energy exchange between reaction coordinate and barrier heat bath is faster as compared to the corresponding long memory time with the same friction magnitude; this means that for high-friction Markovian barrier dynamics, a larger fraction of particles entering the barrier region (without enough energy to cross the barrier top) are able to obtain the missing energy from the barrier heat bath, which leads to a decrease in the \ac{MFPT}.
Interestingly, as can be seen in fig.~\ref{barrierResults_T}A,
this effect changes the \ac{MFPT} in the opposite way as the `non-Markovian-barrier acceleration' predicted by GH theory, and reproduced by our numerical model in the limit of high-friction Markovian dynamics in the well.

Nevertheless, the mechanisms behind both regimes are similar and can each be understood from the committor analysis shown in fig.~\ref{barrierResults_T}F. `Non-Markovian-barrier acceleration` profits from the fact that in case of long barrier memory fast initial velocities, i.e. $v_{\rm in} \gtrsim \vthres$ (to the right of the vertical black broken line in fig.~\ref{barrierResults_T}F), always lead to a direct transition.
In case of short well memory, the initial velocities $v_{\rm in}$ of subsequent barrier crossing attempts are 
only weakly correlated,
 allowing for the assumption made by GH theory that the equilibrium velocity distribution is sampled equally at any attempt. Consequently the high-velocity tail of the equilibrium distribution of $v_{\rm in}$ is visited more frequently over time compared to the long well memory case, where the initial velocity changes rather slowly for consecutive attempts, due to the weak coupling to the well heat bath.
On the other hand, in the limit of long well memory the `Markovian-barrier acceleration` profits from the fact, that in case of slow initial velocities, i.e.~to the left of the vertical black broken line, and short barrier memory, there is still a small chance that a transition over the barrier occurs. This leads to a slightly faster \ac{MFPT} when compared to the case that all the energy to reach the barrier top needs to be accumulated from the well heat bath.

This analysis furthermore suggests a simple way to quantify 
 the `Markovian-barrier acceleration`: 
The global \ac{MFPT} is mainly determined by the \ac{MFPT} to reach the barrier region for the first time. Subsequently, a successful transition of the barrier region happens relatively quickly. 
On the contrary, for long barrier memory the global \ac{MFPT} is determined by the time to reach the barrier top. In fig.~\ref{barrierResults_T}A we therefore also compare the \ac{MFPT} to reach the barrier entry at $x/L=-0.5$ (red and blue dash-dotted lines) and the \ac{MFPT} to cross the barrier at $x/L=0.5$ (data coincides with the red and blue solid lines), both evaluated from simulations with a global memory friction assuming the well friction parameters. 
 As expected, the former coincide with the data of the 'Markovian-barrier accleration' regime correctly, while the later coincide with the data in case of long barrier memory.

A comparison of all presented simulation data with the global analytical rate theory for local memory effects by \citet{Krishnan1992} is shown in appendix \ref{sectionFurtherNumData}\ref{sectionKSR}. Their theory performs well in some regimes of the parameter space, correctly interpolates between predictions by GH and PGH and therefore intrinsically determines whether the dynamics are well or barrier dominated. However, in certain parameter regimes, including the 'Markovian-barrier acceleration' regime, major deviations from the numerical results are observed. 
This is due to instabilities of the perturbation theory inherent to the analytical approach and the authors themselves state that predictions in this parameter regime should be validated by simulations, as we have finally done here.

\section{Conclusions}
\label{disc}

We study a model for barrier crossing with different well and barrier memory friction times and magnitudes. 
By comparing extensive numerical simulations of this model to GH theory (which takes into account memory friction in the barrier region) and PGH theory (which does not take into account space-inhomogeneous memory), 
we identify in which region of the model parameter space the barrier-crossing time, in terms of the \acl{MFPT} (\acs{MFPT}), is 
determined by the well memory or the barrier memory, respectively.

\begin{table}[tb]
\caption{Summary of the regimes observed
when varying single-exponential barrier and well memory friction,
and the respective applicable rate theories with the dominant pre-exponential scaling factors. The table is approximately valid while the barrier and well friction magnitudes remain within one order of magnitude. 
Some effects for very different friction magnitudes in well and barrier are discussed in appendix \ref{sectionFurtherNumData}\ref{sectionInertialBarrier}. }
\setlength\arrayrulewidth{1pt}
\setlength\extrarowheight{1pt}
\setlength{\tabcolsep}{0pt}
\definecolor{b}{rgb}{ 0.5568627451, 0.72549019608, 0.84705882353}
\definecolor{lb}{rgb}{0.7, 0.9, 0.9}
\definecolor{o}{rgb}{1.0, 0.74117647059, 0.52156862745}
\begin{center}
{
\small
\begin{tabular}{@{} c c | c | c | c | @{}}
 &  & \multicolumn{3}{c |}{\bf well} \\ 
 &  & Markovian & Markovian & long \\
 &  & \ high friction\ \ & \ low friction\ \  & memory \\ \hline
\parbox[t]{5mm}{\multirow{14}{*}{\rotatebox[origin=c]{90}{\bf barrier}}} &  
\parbox[t]{9mm}{\multirow{5}{*}{\rotatebox{90}{\parbox{2cm}{Markovian\\ high friction}}}} & \cellcolor{o} barrier dom. &\cellcolor{b} & \cellcolor{lb} well dom. \\ 
 & & \cellcolor{o} Kramers, GH, & \cellcolor{b}   & \cellcolor{lb} PGH  \\ 
 &  & \cellcolor{o} PGH & \cellcolor{b} & \cellcolor{lb} (Markov. \\ 
 &  & \cellcolor{o} MM & \cellcolor{b}  & \cellcolor{lb} barr. acc.) \\ 
  &  & \cellcolor{o} $\tmfp \sim \gB$ & \cellcolor{b} well dom. & \cellcolor{lb} $\tmfp \sim \tW^2/\gW$ \\ \hhline{~-|-|>{\arrayrulecolor{b}}->{\arrayrulecolor{black}}|-|}
 & \multirow{5}{*}{\rotatebox{90}{\parbox{2cm}{Markovian\\ low friction}}}  & \cellcolor{o} barrier dom. & \cellcolor{b} PGH & \cellcolor{b}  \\ 
 & & \cellcolor{o} PGH & \cellcolor{b} MM & \cellcolor{b} \\ 
 &  & \cellcolor{o} MM & \cellcolor{b} $\tmfp$ & \cellcolor{b} \\
  &  & \cellcolor{o} $\tmfp \sim m/\gB$ & \cellcolor{b} $\sim m/\gW$ & \cellcolor{b} well dom.\\
&  & \cellcolor{o}  & \cellcolor{b}  & \cellcolor{b} PGH\\  
   \hhline{~-|-|>{\arrayrulecolor{b}}->{\arrayrulecolor{black}}|>{\arrayrulecolor{b}}->{\arrayrulecolor{black}}|}
 & \multirow{4}{*}{\rotatebox{90}{\parbox{1.5cm}{long\\ memory}}}  & \cellcolor{o} barrier dom. & \cellcolor{b} & \cellcolor{b} $\tmfp \sim \tW^2/\gW$ \\ 
 &  & \cellcolor{o} GH (non-Markov. & \cellcolor{b} & \cellcolor{b} \\ 
  &  & \cellcolor{o} barr. acc.) & \cellcolor{b} & \cellcolor{b} \\
  &  & \cellcolor{o} $\tmfp \sim \lambda^{-1}(\gB,\tB)$ & \cellcolor{b} & \cellcolor{b} \\ \hline
\end{tabular}
}
\end{center}
\label{table}
\end{table} 

The memory friction around the barrier top determines the
\ac{MFPT}
only if the dynamics in the well is in the Markovian high-friction regime.
In this case the \ac{MFPT} is well-described by GH theory if non-Markovian effects on the barrier are present, and instead by MM theory if Markovian low-friction effects dominate the barrier dynamics while the friction magnitudes in well and barrier are comparable.

If the dynamics in the well is in the so-called energy-diffusion regime,
i.e.~either dominated by inertia effects, $\tmW/\tdW \gtrsim 1$,
or because of long memory in the well, $\tW/\tdW \gtrsim 1$, then the rate-limiting step is obtaining enough energy from
the well heat bath to make a barrier-crossing attempt.
In this scenario, the \ac{MFPT} is described by the PGH theory evaluated for the well parameters.
In this regime, high friction
 on the barrier top slightly diminishes the \ac{MFPT}, which to date is not described by
 any rate theory;
this `Markovian-barrier acceleration' is due to the strong interaction between
 reaction coordinate and heat bath in the barrier region, which
enables particles that enter the barrier region without enough energy for a barrier crossing to gain the missing energy in the barrier region and
make it over the top. 
Interestingly, the same mechanism leads to a slow down in case of a Markovian well, where the `non-Markovian-barrier acceleration'  correctly predicted by GH theory happens in the limit of long barrier memory, not short barrier memory. This contrast highlights the complex interplay between barrier friction  and well friction.
The different regimes, and which theory needs to be evaluated where to describe the
corresponding \acp{MFPT}, are summarized in table~\ref{table}.
The table allows to quickly infer which aspect of the dynamics of a reaction
coordinate determines the time scales of rare events,
and will help researchers identify the appropriate rate theory for a
given system.

While theoretical works often only incorporate either space-inhomogeneous friction magnitudes or homogeneous time-dependent memory friction  \cite{Hummer2004,
Hinczewski2010,Berezhkovskii2011,
Pollak1989,Kappler2018,
Daldrop2018,
Kappler2019b,
Kappler2019,
Satija2019,
Lickert2020,
Lavacchi2020},
reaction coordinates in physical systems with non-linear interactions may general exhibit both effects simultaneously.
Our model system therefore represents a step towards more realistic coarse-grained descriptions
of reaction coordinates.
To parametrize a GLE with both space-inhomogeneous memory friction time and magnitude, such as the one presented in this work, from time series data,
an extension of methods established for homogeneous memory can be considered \cite{Daldrop2018, Ayaz2022}.
Furthermore, there are several relevant extensions of our model system.
First, an interaction between the different coupling heat baths could be included, as in a 
physical system the orthogonal degrees of freedom are in general not isolated from each other.
Second, it will be interesting to consider the non-equilibrium scenario where the interaction between reaction coordinate and orthogonal degrees of freedom does not 
originate from an interaction potential; this scenario has been studied before
for homogeneous friction
 \cite{Loos2020}.

Quantum effects are not incorporated in the present model, but projection methods in quantum systems have previously been discussed \cite{Nakajima1958, Ford1988}. 
Under the Born-Oppenheimer approximation classical barrier crossing dynamics would essentially be modified by two effects: reduction of the effective barrier height due to zero-point motion and competition of the classical barrier-crossing rate with the tunneling rate \cite{Chandler1986, Hanggi1990}.
Beyond the Born-Oppenheimer approximation, nonadiabatic effects such as electronic transitions between different energy surfaces would require multistate modeling \cite{Carmeli1985}.

\acknowledgments{
We gratefully acknowledge support by the Deutsche Forschungsgemeinschaft (DFG) grant SFB 1078, by the European Research Council under the Horizon 2020 Programme, ERC grant agreement number 740269, by the Royal Society through grant RP1700 and computing time on the HPC cluster at ZEDAT, FU Berlin.
}

\subsection*{Author contributions}
F.N.B., R.R.N. and J.K. conceived the theory and designed the simulations. F.N.B. and J.K. performed simulations. F.N.B. analyzed the data. All authors discussed the results, analyses and interpretations. F.N.B. and J.K. wrote the paper with input from all authors.

\subsection*{Competing interests}
The authors declare no competing interests.

\clearpage

\appendix

\section{Generalized Langevin equation with space-inhomogeneous memory friction}
\label{sectionDerivations}

\subsection{Formulation of the GLE in a Markovian embedding}
\label{subs_gleDeriv}

In the present section we show that the GLE with space-inhomogeneous memory, eq.~\eqref{eq:gle_local}
from the main text,
is equivalent to a $N+1$-dimensional dynamical system, in which the reaction coordinate
$x(t)$ is coupled to $N$ auxiliary degrees of freedom $(y_1(t),...,y_N(t))$, which we also refer to as the heat
bath. 
We assume that each of the $y_i$ obeys an 
overdamped Langevin equation with random force $F_i(t)$ and friction magnitude $\gamma_i$.
Analogous to the derivation by \citet{Zwanzig1973},
we assume that the reaction coordinate is coupled to the heat bath via 
a
 non-linear potential $U_{\mathrm{hb}}(x,y_1,...y_N)=\sum_{i=1}^N k_i(f_i(x)-y_i)^2/2$,
 where hb stands for heat bath,
the $k_i$ determines the coupling strength between $x$ and $y_i$, 
and the functions $f_i$ will be used to obtain a space-inhomogeneous coupling between reaction coordinate and reservoir $i$.
The total potential $U_{\mathrm{tot}}$ experienced by the dynamical system $(x(t),y_1(t),...,y_N(t))$
is then given as a sum
\begin{equation}
U_{\mathrm{tot}}(x,y_1,...,y_N) = U(x) + U_{\mathrm{hb}}(x,y_1,...,y_N),
\end{equation}
where $U(x)$ is the double well potential eq.~\eqref{eq:double_well_potential} from the main text.
The equations of motion for $x(t)$ and the $y_i(t)$ are then given by
\begin{align}
\label{eq:ddv}
m \ddot x (t) &= - \sum_{i=1}^N k_i \left[ f_i[x(t)]-y_i(t) \right] \ddx f_i[x(t)]  \\ 
\nonumber
&\quad - (\ddx U)[x(t)]
\\
\label{eq:ddw}
\gamma_i \dot y_i(t) &=  k_i \left[ f_i[x(t)] - y_i(t) \right] + F_i(t).
\end{align}
The random forces $F_i$ are Gaussian white noise with
 zero mean, $\langle F_i(t)\rangle=0$, and covariances $\langle F_i(t)F_j(t') \rangle=2 \gamma_i k_BT \delta_{ij} \delta(t-t')$, so that the Langevin eq.~\eqref{eq:ddw} obeys the fluctuation-dissipation relation. 
To obtain a GLE for only the reaction coordinate $x(t)$, we now eliminate the explicitly appearing $y_i(t)$ in 
eq.~\eqref{eq:ddv}.
For this, we use the formal solution of 
 eq.~\eqref{eq:ddw}, which is given by
\begin{align}
\nonumber
y_i(t) &= y_i(0) e^{-t/\tau_i} + \tau_i^{-1} \int_0^{t} dt'\ e^{-(t-t')/\tau_i} f_i[x(t')]\\
&\ \ \ \ + \int_0^{t} dt'\ e^{-(t-t')/\tau_i}\frac{F_i(t)}{\gamma_i} 
\label{eq:ddw_sol}
\\
&= \left[ y_i(0) - f_i[x(0)] \right] e^{-t/\tau_i} + f_i[x(t)]\\
\nonumber
&\ \ \ \ - \int_0^{t} dt'\ e^{-(t-t')/\tau_i} \ddx f_i[x(t')] \dot x(t')\\
\nonumber
&\ \ \ \ + \int_0^{t} dt'\ e^{-(t-t')/\tau_i}\frac{F_i(t)}{\gamma_i},
\end{align}
where we define the relaxation time of reservoir $i$ as $\ti={\gamma_i}/{k_i}$.
Substituting the formal solution for $y_i(t)$ into eq.~\eqref{eq:ddv}, we obtain
\begin{align}
\label{eq:methods_gle}
\begin{split}
m \ddot x (t) = &-  \int_{0}^{t} \Gamma[t-t',x(t),x(t')] \dot x(t') dt'  \\
&- \ddx U[x(t)] + \eta[x(t),t]
\end{split}
\end{align}
with the space-inhomogeneous memory function
\begin{align}
\label{eq:pos_kernel}
\begin{split}
 \Gamma[t-t',x(t),x(t')] &= \\
 \sum_{i=1}^N \frac{\gamma_i}{\ti} \ddx f_i[x(t)] &e^{(t-t')/\ti} \ddx f_i[x(t')],
\end{split}
\end{align}
 and the random force
\begin{align}
\label{eq:pos_noise}
 \eta[x(t),t] &= -\sum_{i=1}^N \frac{\gamma_i}{\ti} \ddx f_i[x(t)] e^{-t/\ti} \left[f_i[x(0)]-y_i(0)\right] \\
 &\qquad+ \sum_{i=1}^N \frac{1}{\ti} \int_0^t dt'\ \ddx f_i[x(t)] e^{(t-t')/\ti} F_i(t').
 \nonumber
\end{align}

How the coupling of the reaction coordinate to reservoir $i$ varies with $x(t)$ is determined 
by the function $f_i(x(t))$.
To obtain an on/off coupling depending on the value of $x(t)$, as used in
 eqs.~\eqref{eq:kernel_local} and \eqref{eq:chi},
 we choose functions
\begin{align}
f_i(x) = 
\begin{cases}
x, &x\in X_i \\ 
\text{min}(X_i), &x < \text{min}(X_i) \\
\text{max}(X_i), &x \geq \text{max}(X_i),
\end{cases},
\end{align}
where $X_i$ is a spatial domain, which we assume to be a single interval,
 within which $x(t)$ couples to reservoir $i$.
With this definition, the spatial derivative of $f_i$ is the coupling function,
\begin{align}
(\ddx f_i)(x) = \chi_i(x) := \begin{cases}1, &x\in X_i \\ 0, &x \notin X_i \end{cases},
\end{align}
so that eqs.~\eqref{eq:ddv} and \eqref{eq:ddw} couple $x(t)$ and $y_i(t)$ if and only if $x(t)\in X_i$, 
equivalent to a local memory kernel in that regime, c.f.~eq.~\eqref{eq:pos_kernel}.
Note that, strictly speaking, 
 the derivative $\partial_x f_i$ is not defined
at the two values, $x = \min(X_i)$, $\max(X_i)$; since the probability that the reaction coordinate takes
either one of these values  is zero, this is not an issue.

To simulate the \ac{GLE} eq.~\eqref{eq:gle_local}, we always use the equivalent formulation
in terms of a dimensionless version of the Markovian system of eqs.~\eqref{eq:ddv}, \eqref{eq:ddw},
given in appendix~\ref{sectionDerivations}\ref{subs_gleDL}.

\subsection{Generalized fluctuation-dissipation relation}
\label{subs_fdt}

We now show that the memory kernel eq.~\eqref{eq:pos_kernel} and the
random force eq.~\eqref{eq:pos_noise} obey the generalized fluctuation-dissipation theorem
\begin{align}
\label{eq:GFTD}
\langle \noiseLoc \eta[x(t'),t'] \rangle&=k_BT\,\Gamma[x(t),x(t'),t-t'].
\end{align}
To compute the autocorrelation on the left-hand side of eq.~\eqref{eq:GFTD} for
 all times $t$, $t'$, and not just for times larger than the longest initial relaxation time $\max_i \{\tau_i\}$ of the heat bath,
  we need to specify initial conditions $y_i(0)$ for the auxiliary variables,
 which appear in eq.~\eqref{eq:pos_noise}.
For this we assume that, for given $x(0)$, the $y_i(0)$ are distributed according to the
  Boltzmann distribution pertaining to the potential $U_{\mathrm{hb}}$, so that $y_i(0) - f_i[x(0)]$ 
  are given by a Gaussian distribution with 
  zero mean and variance $\langle \left[y_i(0) - f_i[x(0)]\right]^2 \rangle = k_BT {\tau_i}/{\gamma_i}$.
  With this initial condition, the autocorrelation of the noise $\noiseLoc$ follows as
\onecolumngrid
\begin{align}
\langle  \eta[x(t),t]  \eta[x(t'),t']\rangle &= \sum_{i=1}^N \left(\frac{\gamma_i}{\ti}\right)^2 \chi_i[x(t)] \chi_i[x(t')] e^{-(t+t')/\ti} \langle \left[f_i[x(0)]-y_i(0)\right]^2 \rangle \nonumber\\
 &\ \ + \sum_{i=1}^N \frac{1}{\ti^2} \int_0^t dt''\ \int_0^{t'} dt'''\ \chi_i[x(t)] \chi_i[x(t')] e^{-(t-t''+t'-t''')/\ti} \langle F_i(t'') F_i(t''')\rangle. \nonumber\\
 &= \sum_{i=1}^N k_BT\frac{\gamma_i}{\ti} \chi_i[x(t)] \chi_i[x(t')] e^{-(t+t')/\ti} 
 \nonumber \\
 &\ \ 
 + \sum_{i=1}^N \frac{1}{\ti^2} \chi_i[x(t)] \chi_i[x(t')] e^{-(t+t')/\ti} \int_0^t dt''\ \int_0^{t'} dt'''\  e^{(t''+t''')/\ti} 2 \gamma_i k_BT \delta(t''-t''') \nonumber \\
 &= \sum_{i=1}^N k_BT\frac{\gamma_i}{\ti} \chi_i[x(t)] \chi_i[x(t')] e^{-(t+t')/\ti} \nonumber \\
 &\ \ + \sum_{i=1}^N 2 k_BT \frac{\gamma_i}{\ti^2} \chi_i[x(t)] \chi_i[x(t')] e^{-(t+t')/\ti} \int_0^{\text{min}(t,t')} dt''\ e^{2t''/\ti}\nonumber \\
 \label{eq:pos_noise_corr}
 &= \sum_{i=1}^N k_BT \frac{\gamma_i}{\ti} \chi_i[x(t)] \chi_i[x(t')] e^{-|t-t'|/\ti}.
\end{align}
\twocolumngrid
By comparing the result eq.~\eqref{eq:pos_noise_corr} with the memory kernel \eqref{eq:pos_kernel},
we observe that the generalized fluctuation-dissipation relation eq.~\eqref{eq:GFTD} holds.

\subsection{Dimensionless formulation of the GLE}
\label{subs_gleDL}

In the present section, we give the dimensionless version of both
 the \ac{GLE} eq.~\eqref{eq:gle_local}, 
 as well as the equivalent Markovian system eqs.~\eqref{eq:ddv} and \eqref{eq:ddw}.
 This in particular makes explicit how many independent parameters the GLE model has.

Using the typical length scale $L$ of the potential eq.~\eqref{eq:double_well_potential},
and the thermal energy $k_BT \equiv \beta^{-1}$ as energy scale,
we define the diffusive time, $\td=\beta L^2\gamma$,
which is the typical time a freely diffusing particle needs to travel a distance $L$ in a 
flat potential landscape, and with friction constant $\gamma = \sum_i \gamma_i$.
We furthermore define the inertial time scale $\tm=m/\gamma$,
on which inertia is dissipated.

Using the scales $L$, $\td$, $\tm$, $\beta$, 
we rewrite the coupled Langevin eqs.~\eqref{eq:ddv} and \eqref{eq:ddw}
in dimensionless form as
\begin{align}
\frac{\tm}{\td} \ddot{\tilde{x}}(\tDL) &= - \sum_{i=1}^N \frac{\gamma_i}{\gamma} \frac{\td}{\tau_i} \left[ \tilde f_i[\xDL(\tDL)]-\yDL_i(\tDL) \right] \ddxDL \tilde f_i[\xDL(\tDL)] \nonumber \\
\label{eq:ddvDL}
 &\quad - (\ddxDL \UDL)[\xDL(\tDL)]\\
\label{eq:ddwDL}
\dot \yDL_i(\tDL) &=   \frac{\td}{\tau_i} \left[ \tilde f_i[\xDL(\tDL)] - \yDL_i(\tDL) \right] + \sqrt{\frac{\gamma}{\gamma_i}} \FDL_i(\tDL),
\end{align}
where $\tDL := t/\td$, $\xDL(\tDL) := x(\td \tDL)/L$, $\dot{\tilde{x}}(\tDL) = \td \dot{x}(\td \tDL)/L$, $\ddot{\tilde{x}}(\tDL) = \td^2 \ddot{x}(\td \tDL)/L$ are dimensionless time, position, 
velocity and acceleration. 
$-(\ddxDL \UDL)(\xDL)$ 
is the dimensionless deterministic force corresponding to the quartic potential 
eq.~\eqref{eq:double_well_potential}, 
and given by
\begin{align}
-(\ddxDL \UDL)(\xDL) = - 4 \UDL_0 \left(\xDL^2-1\right) \xDL.
\end{align}
with dimensionless barrier height $\UDL_0 := \beta U_0$. 

The dimensionless coupling between reaction coordinate and heat bath
is given by $\tilde{f}_i(\xDL) := f_i(L \xDL)/L$, so that
\begin{align}
\tilde f_i(\xDL) = 
\begin{cases}
\xDL, &\xDL \in \XDL_i \\ 
\text{min}(\XDL_i), &\xDL < \text{min}(\XDL_i) \\
\text{max}(\XDL_i), &\xDL \geq \text{max}(\XDL_i)
\end{cases},
\end{align}
where $\XDL_i = X_i/L$, so that 
\begin{align}
(\ddxDL \fDL_i)(\xDL) = \widetilde \chi_i(\xDL) = \begin{cases}1, &\xDL \in \XDL_i, \\ 0, 
&\xDL \notin \XDL_i.\end{cases}
\end{align}

The dimensionless random forces  $\FDL_i(\tDL) = L/(k_BT) \sqrt{\gamma/\gamma_i} F_i(t)$ are 
Gaussian white noise with zero mean, 
and covariances 
$\langle \FDL_i(\tDL)\FDL_j(\tDL') \rangle 
= 2 \delta_{ij} \delta(\tDL-\tDL')$.

As in sect.~\ref{subs_gleDeriv}, we formally solve eq.~\eqref{eq:ddwDL}, and substitute the 
result into eq.~\eqref{eq:ddvDL}, to obtain the dimensionless GLE
\begin{align}
\label{eq:gleTD}
\begin{split}
\frac{\tm}{\td}\ddot{\tilde{x}}(\tDL) = &- \int_{0}^{\tDL} \GammaDL[\tDL-\tDL',\xDL(\tDL),\xDL(\tDL-\tDL')] \dot{\tilde{x}}(\tDL')\,\mathrm{d}\tDL' \\
  &-(\ddxDL \UDL) \left[\xDL(\tDL)\right] + \noiseLocDL,
\end{split}
\end{align}
with the dimensionless space-inhomogeneous memory kernel
\begin{align}
\label{eq:pos_kernel_dl}
\begin{split}
\GammaDL[\tDL-\tDL',\xDL(\tDL),\xDL(\tDL')] &=  \\
\sum_{i=1}^N \frac{\gamma_i}{\gamma} \frac{\td}{\ti} \widetilde \chi_i[\xDL(\tDL)] \widetilde \chi_i[ \xDL(\tDL')]
 &\exp\left[-\frac{\td}{\ti}\left(\tDL-\tDL'\right)\right],
\end{split}
\end{align}
and the dimensionless random force $\noiseLocDL:=\beta \eta[x(t),t]/L$.
Instead of carrying out the projection method again, as we did in the present section, 
eqs.~\eqref{eq:gleTD}, \eqref{eq:pos_kernel_dl},
can also be obtained by directly recasting eqs.~\eqref{eq:methods_gle}, \eqref{eq:pos_kernel},
in dimensionless form.
Similarly to eq.~\eqref{eq:GFTD}, 
the dimensionless memory kernel $\GammaDL$ and random force $\tilde{\eta}$
obey the generalized fluctuation-dissipation theorem
 \begin{align}
\label{eq:dl_GFDT}
\begin{split}
\langle &\noiseLocDL \tilde \eta[\xDL(\tDL'),\tDL']\rangle
= \GammaDL[\tDL-\tDL',\xDL(\tDL),\xDL(\tDL')].
\end{split}
\end{align}

\section{Rate theories}
\label{rateTheorySection}

\subsection{Formulas for rate theories considered in the main text}

In the present section we recall the formulas used to evaluate the various
classical rate theories we consider in the main text.

\subsubsection{Transition-state theory}
While we do not explicitly show results from transition-state theory (TST) \cite{Eyring1935} in the main text,
the TST escape rate appears in several of the rate theories we consider.
According to TST, for a parabolic free-energy in the reactant state, 
the mean escape time is given as \cite{Eyring1935}
\begin{align}
\label{eq:tst}
\tau_{\mathrm{TST}} = \frac{2\pi}{\OmegaMin} e^{\beta U_0},
\end{align}
where as before $U_0$ denotes the barrier height, $\beta^{-1}=k_BT$ is the thermal energy,
 and the well frequency $\OmegaMin = \sqrt{U''_{\mathrm{min}}/m} $ 
 contains the curvature $U''_{\mathrm{min}} := U''(x_{\mathrm{min}})$ at the minimum $x_{\mathrm{min}}$
  of the potential well from which the particle escapes. 

\subsubsection{Kramers' theory}

Kramers considered the escape from a potential well for a particle described by
 the Markovian inertial Langevin equation,
for both the limits of medium-to-high friction, and low friction \cite{Kramers1940}.
For the medium-to-high friction regime, he obtained
\begin{align}
\label{eq:kramers}
 \tau_{\mathrm{Kr}}^{\mathrm{hf}} =  \left[ \left( \frac{\gamma^2}{4m^2} +\OmegaMax^2 \right)^{1/2}-\frac{\gamma}{2m}\right]^{-1} \OmegaMax \tau_{\mathrm{TST}},
 \end{align}
while in the low-friction limit, he derived
 \begin{align}
\label{eq:kramers_inertial}
 \tau_{\mathrm{Kr}}^{\mathrm{lf}} = \frac{m}{\gamma \beta U_0} e^{\beta U_0},
 \end{align}
where the barrier frequency $\OmegaMax = \sqrt{U''_{\mathrm{max}}/m} $ contains the curvature $U''_{\mathrm{max}} := U''(x_{\mathrm{max}})$ at
 the barrier top $x_{\mathrm{max}}$. 
 Note the opposite scaling of both equations with respect to the friction constant $\gamma$: 
 While for medium-to-high friction it holds that $\tau_{\mathrm{Kr}}^{\mathrm{hf}}\sim \gamma$, 
 for low friction we have $\tau_{\mathrm{Kr}}^{\mathrm{lf}}\sim \gamma^{-1}$.

\subsubsection{Mel'nikov and Meshkov theory}

\citet{Melnikov1986} (MM) derived a solution to the Kramers' problem
which is valid for all values of the friction,
and hence bridges the two asymptotic expressions 
eqs.~\eqref{eq:kramers}, \eqref{eq:kramers_inertial}.
The MM result is given by
\begin{align}
\label{eq:MM}
 \tau_{\mathrm{MM}} &=  A(\Delta) \left[ \left( \frac{\gamma^2}{4m^2} +\OmegaMax^2 \right)^{1/2}-\frac{\gamma}{2m}\right]^{-1} \OmegaMax \tau_{\mathrm{TST}}, \\
 A(\Delta) &= \exp\left[\frac{2}{\pi}\int_{0}^{\frac{\pi}{2}}\ln\left[1-e^{\Delta/4 \cos^2(x)}\right]dx\right], \\
 \Delta &=  2  \sqrt{2} \frac{\gamma}{\sqrt{m}} \beta \int_{-\sqrt{2}L}^0 \sqrt{U_0-U(x)}\ dx.
\end{align}

\subsubsection{\acl{GH} theory}

While both Kramers' and \acl{MM} theory consider Markovian dynamics,
 \citet{Grote1980} developed a theory for \acl{MFPT} under influence of memory effects. 
 Their expression for the case where the dynamics in the potential wall relax fast, 
 and only memory effects on the barrier are relevant, is given by
\begin{align}
\label{eq:ght}
  \tau_{\mathrm{GH}} &= \frac{\OmegaMax}{\lambda} \tau_{\mathrm{TST}},
\end{align}
where $\tilde{\Gamma}(\lambda)$ denotes the Laplace-transformed memory friction kernel $\Gamma(t)$
at the barrier top,
and the reactive frequency $\lambda>0$ is given as the solution of the equation
\begin{align}
  \label{eq:gh_lambda}
  \lambda &= \frac{\OmegaMax^2}{\lambda + \tilde{\Gamma}(\lambda) / m }.
\end{align}
Thus, for a single exponential kernel $\Gamma(t) = {\gamma} e^{-t/\tau}/{\tau}$,  
 $\lambda$ is calculated from 
the cubic equation
\begin{align}
\lambda^3 + \frac{\lambda^2}{\tau} + \left(\frac{\gamma}{m\tau}- \OmegaMax^2 \right) \lambda = \frac{\OmegaMax^2}{\tau}.
\end{align}

Note that, either in the inertial, $m\to \infty$, or the long memory limit, $\tau\to \infty$, it follows $\lambda=\OmegaMax$ and GH theory collapses onto the transition-state theory result, $\tau_{\mathrm{GH}}=\tau_{\mathrm{TST}}$ in eq.~\eqref{eq:ght}.

Furthermore, in case of instantaneous, i.e. delta-correlated friction, $\Gamma(t) = {\gamma} \delta(t)$ and $\tilde \Gamma(\lambda) = {\gamma}$,  it follows $\lambda= \left( \gamma^2/(4m^2) +\OmegaMax^2 \right)^{1/2}-\gamma/(2m)$, which results in $\tau_{\mathrm{GH}}=\tau_{\mathrm{Kr}}^{\mathrm{hf}}$, the Kramers high-friction result in eq.~\eqref{eq:kramers}.

\subsubsection{Heuristic formula}

The heuristic formula was fitted to agree with both the theory by \citet{Pollak1989}
and 
numerical simulations of the \ac{MFPT} in the double-well potential eq.~\eqref{eq:double_well_potential},
and a GLE
with a global single-exponential memory kernel 
with friction magnitude $\gamma$ and memory time $\tau$ \cite{Kappler2018, Kappler2019}. 
Using the diffusive and inertial time scales $\td=\gamma \beta L^2$ and $\tm=m/\gamma$,
the heuristic formula is given by 
\begin{align}
\frac{\tau_{\text{emp.}}}{\td} &= \frac{e^{\beta U_0}}{\beta U_0}
\left[ \frac{\pi}{2\sqrt{2}} \left( 1+ 10 \beta U_0 \frac{\tau}{\td} \right)^{-1}+\frac{\tm}{\td} 
\right.
\\ \nonumber
& \qquad\qquad\qquad \left.
+ 2 \sqrt{\beta U_0 \frac{\tm}{\td}}+ 4 \beta U_0 \frac{\tau ^2}{\td^2} \right].
\end{align}

\subsection{Evaluation of rate theories for a space-inhomogeneous memory kernel}
\label{effParamSection}

Whenever we evaluate a rate theory for the effective friction parameters of a region, we use the respective regional friction- and memory parameters, or equivalently $\tau_i$, $\tau_{D,i} = L^2 \beta \gamma_i$, $\tau_{m,i} = m/\gamma_i$. 
On the other hand, in plots we always rescale the MFPT as well as the parameters using the
diffusive- and inertial times $\td = L^2 \beta \gamma$, $\tm = m/\gamma$, which correspond to the total friction magnitude $\gamma = \sum_i \gamma_i$.
 In the present section we state the relevant relations between these local and global 
 timescales.

The relation between the local and global diffusive- and inertial time scales is given by
\begin{align}
    \tau_{D,i} &= L^2 \beta \gamma_i = \frac{\gamma_i}{\gamma} L^2 \beta \gamma =  \frac{\gamma_i}{\gamma} \tau_D,\\
    \tau_{m,i} &= \frac{m}{\gamma_i} = \frac{\gamma}{\gamma_i} \frac{m}{\gamma} = \frac{\gamma}{\gamma_i} \tau_m,
\end{align}
so that
\begin{align}
    \frac{\tau_{m,i}}{\tau_{D,i}} &= \left( \frac{\gamma}{\gamma_i}\right)^2 \frac{\tau_m}{\tau_D},\\
    \frac{\tau_i}{\tau_{D,i}} &= \frac{\gamma}{\gamma_i} \frac{\tau_i}{\tau_D},\\
    \frac{\tau_{\mathrm{MFP}}}{\tau_{D,i}} &= \frac{\gamma}{\gamma_i} \frac{\tau_{\mathrm{MFP}}}{\tau_{D}}.
\end{align}
Therefore, if we want to calculate $\tau_{\mathrm{MFP}}/\tau_{D}$ for region $i$ using a rate theory for globally homogeneous friction, we have to evaluate
\begin{equation}
    \frac{\tau_{\mathrm{MFP}}}{\tau_D} = \frac{ \gamma_i}{\gamma} \left.\frac{\tau_{\mathrm{MFP}}}{\tau_{D,i}} \right|_{(\gamma/\gamma_i)^2\tau_m/\tau_D, ~(\gamma/\gamma_i)\tau_i/\tau_D},
\end{equation}
where the first argument $(\gamma/\gamma_i)^2\tau_m/\tau_D$ is the argument for the dimensionless inertial time scale  $\tau_m/\tau_D$ in the rate theory, 
and the second argument $(\gamma/\gamma_i) \tau_i/\tau_D$ is the argument for the dimensionless single-exponential memory $\tau_{\Gamma}/\tau_D$ in the rate theory.

\section{Further comparisons of numerical data with rate theories}
\label{sectionFurtherNumData}

\subsection{Very unequal friction coefficients in well and barrier regions}
\label{sectionInertialBarrier}

\begin{figure}[htb]
  \begin{overpic}[width=0.8\columnwidth]{{/../figs/barrierReset_flo_pub/TM0.001_overGB_log_wGHTbothTau}.pdf}
  \put(0,66){ \huge A}
  \end{overpic}
 \begin{overpic}[width=0.8\columnwidth]{{/../figs/barrierReset_flo_pub/TM0.001_overGW_logbothTau}.pdf}
 \put(0,66){ \huge B}
 \end{overpic}
 \caption{\Acl{MFPT}, $\tmfp/\td$, for various single-exponential barrier- and well- memory friction
 parameters,
compared with analytic predictions given by \citet{Grote1980} (GH, solid line in A), \citet{Melnikov1986} (MM, dotted lines), \citet{Pollak1989} (PGH, evaluated using the heuristic formula \cite{Kappler2019}) and transition-state theory (TST, gray broken lines).
The numerical data is shown for equal barrier and well-friction times, $\tB/\td=\tW/\td=10^{-3}$ and the inertial time scale is fixed in the high-friction regime $\tm/\td=10^{-3}$. Blue square markers denote data for which the local friction-time scales are kept constant, $\tB/\tdB=\tW/\tdW=10^{-3}$, instead of the global ones.
{\bf A:} Results for the limit $\gB/\gamma \ll 1$, i.e.~the limit of Markovian low friction on the barrier.
{\bf B:} Results for the limit $\gW/\gamma \ll 1$, i.e.~the limit of Markovian low friction in the well.
 }
 \label{inertialBarrier}
\end{figure}

In fig.~\ref{barrierResults_G}D in the main text we show both the well- and barrier-evaluated MM predictions
 for high friction, $\tm/\td=10^{-4}$. Both curves show a non-monotonic behavior as a function of
  the barrier friction magnitude $\gamma_{\rm B}/\gamma$:
  Whereas the barrier-evaluated MM theory displays a minimum at small $\gamma_{\rm B}/\gamma$, 
  the well-evaluated MM prediction for the MFPT becomes minimal at $\gamma_{\rm B}/\gamma$ close to 1.
We here  investigate these two limits in detail by performing simulations for 
both $\gamma_{\rm B}/\gamma \ll 1$ and $1 - \gamma_{\rm B}/\gamma = \gamma_{\rm W}/\gamma \ll 1$.
We show the results in fig.~\ref{inertialBarrier}, where we consider an inertial time scale of $\tm/\td=10^{-3}$ 
and equal friction-time scales in the well and barrier, $\tB/\td=\tW/\td=10^{-3}$.

In fig.~\ref{inertialBarrier}A we consider the limit $\gB/\gamma \ll 1$, 
i.e.~the limit of Markovian low friction on the barrier. 
While the MM and PGH theories, both evaluated for the barrier friction parameters, show a non-monotonic trend (namely the Kramers turnover), 
the numerical \ac{MFPT} (orange circles) levels off to a constant for small $\gB/\gamma$, with a value close to the prediction of transition-state theory (gray broken line). 
This limit is also correctly recovered by GH theory evaluated for the barrier friction parameters (orange solid line). Therefore, for $\gB/\gamma \ll 1$, GH outperforms MM theory and PGH theory. 
This is in contrast to the results shown in the main text in fig.~\ref{barrierResults_G}A, D and E, where the local friction magnitudes were considered to be within one order of magnitude, and the well-friction evaluated MM theory described the numerical data.

We discuss the opposite limit $\gW/\gamma \ll  1$ in fig.~\ref{inertialBarrier}B. 
Here the numerical \ac{MFPT} (orange circles) is consistent with
 both MM and PGH theories, 
evaluated on the barrier parameters,
and markedly different from the predictions of transition-state theory and PGH theory evaluated on the well parameters.
The agreement of barrier-evaluated theories and the numerical data indicates that the system is described
by the Kramers high-friction limit.
At first sight this might seem surprising, because for $\gW/\gamma \ll 1$ the local dynamics in the well 
is clearly underdamped, as for $\gW/\gamma=10^{-3}$ we have $\tmW/\tdW = 10^3$.
However, since for $\gW/\gamma \ll 1$ we have $\tdW \ll \tdB$, even though the mean time a particle
in the well needs to reach the barrier is a large multiple of $\tdW$, this time may still be much smaller 
than the time to diffusively cross the barrier (which depends on $\tdB$).
Therefore, for $\gW/\gamma \ll 1$, even though the well dynamics is in the energy diffusion limit,
the crossing over the barrier can still be the rate-limiting step of the escape process.

Furthermore, we note that in both suplots A and B, for the lowest value of the local friction magnitude $\gamma_i/\gamma=10^{-3}$, the local friction times, $\ti/\tdi=1$, are not anymore in the Markovian limit. 
To exclude local non-Markovian effects influencing the shown \acp{MFPT}, we also show numerical data for constant local friction times,  $\tB/\tdB=\tW/\tdW=10^{-3}$ (blue squares) in  fig.~\ref{inertialBarrier}A and B.
This data is almost identical to the data at constant $\tB/\td$, $\tW/\td$,
so that we conclude that non-Markovian effects remain negligible for the parameter regime shown.

\subsection{Variation of barrier height}
\label{sectionBarrierHeight}

\begin{figure}[htb]
  \begin{overpic}[width=0.8\columnwidth]{{/../figs/barrierReset_flo_pub/TM0.000_TB1.000_overU_varTLR}.pdf}
  \put(-10,72){ \huge A}
  \end{overpic}
 \begin{overpic}[width=0.8\columnwidth]{{/../figs/barrierReset_flo_pub/TM0.000_TB0.001_overU_varTLR}.pdf}
 \put(-10,72){ \huge B}
 \end{overpic}
 \caption{\Acl{MFPT}, $\tmfp/\td$, for various single-exponential barrier- and well-friction
 parameters,
shown as a function of the barrier height $\beta U_0$.
The numerical data is shown for two well-friction times, $\tW/\td=10^{-4}$ and $\tW/\td=1$, 
fixed barrier-friction time $\tB/\td=1$ ({\bf{A}}) and $\tB/\td=10^{-3}$ ({\bf{B}}), always for equal friction constants $\gB/\gamma = \gW/\gamma = 1/2$.
 The inertial time scale is fixed in the high-friction regime $\tm/\td=10^{-4}$.
For comparison, predictions by \citet{Grote1980} (GH, broken lines) and \citet{Pollak1989} (PGH, solid lines, evaluated using the heuristic formula \cite{Kappler2019}) are also shown. 
 }
 \label{barrierHeight}
\end{figure}

In the main text, we consider numerical results and rate theories for the 
barrier height $\beta U_0 = 3$.
In fig.~\ref{barrierHeight}, we compare numerical results for barrier heights ranging from $\beta U_0 = 2$
to $\beta U_0 = 7$ to rate-theory predictions, and find that conclusions drawn in the main text remain
true also for the parameters considered here.

While in fig.~\ref{barrierHeight}A we show results for non-Markovian barrier dynamics $\tB/\td = 1$, 
in fig.~\ref{barrierHeight}B we consider a barrier with Markovian dynamics, $\tB/\td = 0.001$.
For both subplots we use $\gW/\gamma=\gB/\gamma=0.5$, i.e.~an equal partitioning of the total friction magnitude to well and barrier, and high friction $\tm/\td=10^{-4}$.
For both subplots, we consider two representative parameters for the well-friction time $\tW/\td$.
The results for $\tW/\td=10^{-4}$, shown as orange circles in fig.~\ref{barrierHeight},
 correspond to the Markovian high-well-friction regime, 
 for which the dynamics are predicted by \ac{GH} theory.
The blue squares in the figure correspond to \acp{MFPT} for well-friction time $\tW/\td=1$, 
which corresponds to the long-well-memory regime
where the \ac{MFPT} is predicted by PGH theory, evaluated using the memory kernel parameters around at the potential well.
As fig.~\ref{barrierHeight}B demonstrates, that the `Markovian-barrier acceleration' regime discussed in detail in the main text also exists for larger barrier heights.
 
Overall, 
fig.~\ref{barrierHeight} shows that the predictions from the main text
 are consistent with the numerical data for all barrier heights $\beta U_0 \in [2,7]$ considered here. 
 In fact the predictions seem to improve for higher barriers. 
 The exponential dependence of the \ac{MFPT} on the barrier-height, 
 known since Arrhenius \citep{Arrhenius1889}, is also well visible in our semi-logarithmic representation
 of the data.

\subsection{Variation of inertial time scale}
\label{sectionBarrierTB}

\begin{figure*}[htb]
 \begin{overpic}[width=0.32\textwidth]{{../figs/barrierReset_flo_pub/TM0.000_overTB_varTLR}.pdf}
 \put(-2,68){\huge A}
 \end{overpic}
 \begin{overpic}[width=0.32\textwidth]{{../figs/barrierReset_flo_pub/TM0.000_overTLR_varTB}.pdf}
 \put(-2,68){\huge B}
 \end{overpic}
\begin{overpic}[width=0.32\textwidth]{{/../figs/barrierReset_flo_pub/map_TM0.000_TLRoverTB}.png}
\put(-2,68){\huge C}
 \end{overpic}

\begin{overpic}[width=0.32\textwidth]{{../figs/barrierReset_flo_pub/TM0.010_overTB_varTLR}.pdf}
\put(-2,68){\huge D}
 \end{overpic}
\begin{overpic}[width=0.32\textwidth]{{../figs/barrierReset_flo_pub/TM0.010_overTLR_varTB}.pdf}
\put(-2,68){\huge E}
 \end{overpic}
\begin{overpic}[width=0.32\textwidth]{{/../figs/barrierReset_flo_pub/map_TM0.010_TLRoverTB}.png}
\put(-2,68){\huge F}
 \end{overpic}

\begin{overpic}[width=0.32\textwidth]{{../figs/barrierReset_flo_pub/TM1.000_overTB_varTLR}.pdf}
\put(-2,68){\huge G}
 \end{overpic}
\begin{overpic}[width=0.32\textwidth]{{../figs/barrierReset_flo_pub/TM1.000_overTLR_varTB}.pdf}
\put(-2,68){\huge H}
 \end{overpic}
   \begin{overpic}[width=0.32\textwidth]{{/../figs/barrierReset_flo_pub/map_TM1.000_TLRoverTB}.png}
\put(-2,68){\huge I} 
 \end{overpic}
 \caption{\Acl{MFPT} (\acs{MFPT}), $\tmfp/\td$, for different barrier memory friction, 
and well memory friction, 
compared with analytic predictions given by \citet{Grote1980} (GH, broken lines) and \citet{Pollak1989} (PGH, solid lines, evaluated using the heuristic formula \cite{Kappler2019}). 
The data is shown for various barrier-friction $\tB/\td$ and well-friction times $\tW/\td$ and equal friction magnitudes $\gB/\gamma = \gW/\gamma = 0.5$. The inertial time scale is constant and different in each row (A--C: $\tm/\td=10^{-4}$, D--F: $\tm/\td=10^{-2}$, G--I: $\tm/\td=1$). {\bf A, D, G:} \ac{MFPT} plotted over the barrier-friction time $\tB/\td$. The theories are shown for the respective barrier-friction time in gray and in case of the PGH theory for the well-friction time as colored solid lines. {\bf B, E, H:} \ac{MFPT} plotted over the well-friction time $\tW/\td$. The theories are shown for the respective well-friction time in gray and in case of the GH for the barrier-friction time as colored broken lines. \textbf{C, F, I:} Contour plots of agreement of the simulation results with the theoretical predictions. The color denotes whether the simulated $\tmfp \in [1/3\,\tau_{\mathrm{theo.}}, 3\,\tau_{\mathrm{theo.}}]$, where $\tau_{\mathrm{theo.}}$ is calculated using either the GH theory with the barrier-friction parameters or the PGH theory with the well-friction parameters. The hatching indicates that both theoretical predictions agree with the simulated data. The light blue area denotes the `Markovian-barrier acceleration' of the PGH prediction for which we define $\tau_{\mathrm{theo.,MBA}}=0.2\,\tau_{\mathrm{theo.,PGH}}$.
}
 \label{barrierResults_overTB_varTBLR}
\end{figure*}

In fig.~\ref{barrierResults_T} of the main text, 
we consider barrier crossing for the high-friction regime $\tm/\td = 10^{-4}$, while the friction magnitudes are equal $\gB/\gamma=\gW/\gamma=0.5$ and the friction times in the well $\tW/\td$ and on the barrier $\tB/\td$ are varied.
In fig.~\ref{barrierResults_overTB_varTBLR}, 
we show additional numerical and analytical \acp{MFPT} for larger inertial times.
We consider the inertial times $\tm/\td=10^{-4}$ (figs.~\ref{barrierResults_overTB_varTBLR}A--C), $10^{-2}$ (figs.~\ref{barrierResults_overTB_varTBLR}D--F), and $1$ (figs.~\ref{barrierResults_overTB_varTBLR}G--I).
While the first column of 
fig.~\ref{barrierResults_overTB_varTBLR} 
shows the rescaled \ac{MFPT} as a function of the barrier-friction time $\tB/\td$ for various
values of the well-friction time $\tW/\td$,
in the second column we vary the well-friction time for several constant values of the barrier-friction time.
In both the first and second column, 
the appropriate analytic predictions including memory effects are given either 
by GH theory, which is evaluated for the effective barrier-friction parameters, 
determined by $\tB$ (broken lines) or by PGH theory (solid lines), 
which is evaluated for the effective well-friction parameters, given by $\tW$. 
The third column of fig.~\ref{barrierResults_overTB_varTBLR} depicts phase diagrams
 that summarizes for which
 parameters $(\tB/\td, \tW/\td)$ the numerical data
agrees with the predictions of GH theory or PGH theory.
Note that fig.~\ref{barrierResults_overTB_varTBLR}A--C are replots of 
 fig.~\ref{barrierResults_T}A,D and E  in the main text. 

Fig.~\ref{barrierResults_overTB_varTBLR} shows that all conclusions drawn in the main text
also hold true as $\tm/\td$ is varied, i.e.~away from the high-friction limit.
In particular, for larger inertial time scales $\tm/\td$, the predictions for the \aclp{MFPT} are described
 globally by PGH theory for the well-friction parameters. 
 Furthermore, figs.~\ref{barrierResults_overTB_varTBLR}D and G show that the `Markovian-barrier acceleration' regime is also present for larger inertial times.
On the other hand, the `non-Markovian-barrier acceleration' predicted by GH theory vanishes.

\subsection{Comparison of numerical results to KSR theory}
\label{sectionKSR}

In the present section, we compare our numerical \acp{MFPT} with the predictions of
 a theory for barrier crossing with space-inhomogeneous memory friction.
\ac{KSR} \cite{Krishnan1992} derived an analytic theory for \acp{MFPT} in a piecewise harmonic potential with different well and barrier memory friction; this theory is based on the formalism by \acl{PGH} \cite{Pollak1989}. 
The analytical \ac{KSR} predictions for the \acp{MFPT} have not yet been compared to numerical simulations. 

The \ac{KSR} model takes as input the memory-friction kernels for the well and barrier regions, for both of which we consider single exponentials, the particle mass $m = \tm \gamma$, the local angular frequencies of the potential for the well, $\omega_0= (\ddx^2 U)(x = -L)/m$ and barrier $\omega_b = -(\ddx^2 U)(x=0)/m$, and the barrier height $\beta U_0 = 3$, where as before $U(x)$ is the quartic potential eq.~\eqref{eq:double_well_potential}.

We now compare the predictions of \ac{KSR} theory with the same numerical data as considered in figs.~\ref{barrierResults_G} and \ref{barrierResults_T} of the main text.
 
First,  in fig.~\ref{KSR_results_G} we consider the data from fig.~\ref{barrierResults_G} of the main text.
For a detailed discussion of the data we refer to the main text, as this section focuses on evaluating the quality of the KSR theory with respect to the other theories. 
For the Markovian limit, i.e.~$\tW/\td = \tB/\td \ll 1$, we generally observe good agreement 
between numerical data and \ac{KSR} theory throughout fig.~\ref{KSR_results_G}.
However, the predictions by \ac{MM} and PGH using effective local parameters perform 
slightly better in the whole parameter range. 
Away from the Markovian limit, for $\tW/\td = \tB/\td = 1$ the predictions by \ac{KSR} disagree considerably with 
the numerical data, as seen for the thick blue line in fig.~\ref{KSR_results_G}C; nonetheless, 
the general trend is predicted correctly also in this regime.
Of course, the true strength of the model by \ac{KSR} here is the correct interpolation between barrier- and well-dominated dynamics,  which needs to be chosen by hand in the evaluation of PGH and MM theories. 
This is most clearly seen in fig.~\ref{barrierResults_G}, where \ac{KSR} theory switches between the barrier-dominated and well-dominated MM predictions as $\tm/\td$ is increased.

\begin{figure*}[htb]
 \begin{overpic}[width=0.32\textwidth]{{/../figs/barrierReset_flo_pub/TBLR0.000_overTM_varGB_KSR_MM}.pdf}
 \put(-2,62){\huge A}
 \end{overpic}
 \begin{overpic}[width=0.32\textwidth]{{/../figs/barrierReset_flo_pub/TBLR0.000_overGB_varTM_KSR}.pdf}
 \put(-2,62){\huge B}
 \end{overpic}
 \caption{\Acl{MFPT} (\acs{MFPT}), $\tmfp/\td$, for different barrier memory friction, $\GB(t)=\gB/\tB e^{-t/\tB}$ and well memory friction, $\GW(t)=\gW/\tW e^{-t/\tW}$, compared with analytic predictions given by \citet{Melnikov1986} (MM) and \citet{Pollak1989} (PGH, evaluated using the heuristic formula \cite{Kappler2019}) as well as the predictions by \citet{Krishnan1992} (KSR, dotted lines), equivalent to figs.~2A and D in the main text. The data is shown for various inertial time scales $\tm/\td$ and equal memory times in the Markovian limit with $\tB/\td=\tW/\td=10^{-4}$. \textbf{A:} \ac{MFPT} plotted over the inertial time scale $\tm/\td$ for different ratios of the barrier friction constant to total friction $\tB/\gamma$. \textbf{B:} \ac{MFPT} plotted over $\tB/\gamma$ for various $\tm/\td$. The predictions by \ac{MM} are shown for the effective barrier-friction parameters, given by $\gB$, as broken lines and for the effective well-friction parameters, given by $\gW$, as solid lines. 
 }
 \label{KSR_results_G}
\end{figure*}

The data of fig.~\ref{barrierResults_T}A and D of the main text is discussed in this section in figs.~\ref{KSR_results_T}A and B,
where $\tm/\td = 10^{-4}$.
While the predictions by \ac{KSR} again interpolate correctly between well- and barrier-dominated dynamics, 
in the regime where $\tW/\td\geq0.1$ and $\tB/\td\leq0.1$, we observe significant deviations
between the numerical \ac{MFPT} and the corresponding \ac{KSR} prediction.
As can be seen clearly in the upper left corner of fig.~\ref{barrierResults_T}A and the right side
of subplot B, the numerical and analytical data can deviate by several orders of magnitude.
For larger inertial time scales, these deviations become smaller
 as shown in figs.~\ref{KSR_results_T}C and D where $\tm/\td=10^{-2}$. 
 We note that \ac{KSR} throughout predicts a barrier crossing speedup as $\tB/\td$ is increased (see subplots A, D), 
 whereas the numerical data displays the `Markovian-barrier acceleration' behavior for $\tW/\td \gtrsim 1$, for which
 barrier crossing is in fact slower as $\tB/\td$ is increased.
Possible explanations for the deviations observed in figs.~\ref{KSR_results_G} and \ref{KSR_results_T} are discussed in the following.

\begin{figure*}[htb]
\begin{overpic}[width=0.32\textwidth]{{../figs/barrierReset_flo_pub/TM0.000_overTB_varTLR_wKSR}.pdf}
 \put(-2,68){\huge A}
 \end{overpic}
 \begin{overpic}[width=0.32\textwidth]{{../figs/barrierReset_flo_pub/TM0.000_overTLR_varTB_KSR}.pdf}
 \put(-2,68){\huge B}
 \end{overpic}
 \begin{overpic}[width=0.32\textwidth]{{../figs/barrierReset_flo_pub/epsDelta_TM0.000_overTB_varTLR}.pdf}
 \put(-2,68){\huge C}
 \end{overpic}
\begin{overpic}[width=0.32\textwidth]{{../figs/barrierReset_flo_pub/TM0.010_overTB_varTLR_wKSR}.pdf}
 \put(-2,68){\huge D}
 \end{overpic}
\begin{overpic}[width=0.32\textwidth]{{../figs/barrierReset_flo_pub/TM0.010_overTLR_varTB_KSR}.pdf}
 \put(-2,68){\huge E}
  \end{overpic}
  \begin{overpic}[width=0.32\textwidth]{{../figs/barrierReset_flo_pub/epsDelta_TM0.010_overTB_varTLR}.pdf}
 \put(-2,68){\huge F} 
 \end{overpic} 
 \caption{\Acl{MFPT} (\acs{MFPT}), $\tmfp/\td$, for different barrier memory friction, $\GB(t)=\gB/\tB e^{-t/\tB}$ and well memory friction, $\GW(t)=\gW/\tW e^{-t/\tW}$, compared with analytic predictions given by \citet{Grote1980} (GH, broken lines) and by \citet{Pollak1989} (PGH, solid lines, evaluated using the heuristic formula \cite{Kappler2019}), as well as predictions by \citet{Krishnan1992} (KSR, dotted lines). The data is shown for various barrier-friction $\tB/\td$ and well-friction times $\tW/\td$ and equal friction constants $\gB/\gamma = \gW/\gamma = 1/2$. The inertial time scale is constant and different in each row (A-C: $\tm/\td=10^{-4}$, D-F: $\tm/\td=10^{-2}$). {\bf A, D:} \ac{MFPT} plotted over the barrier-friction time $\tB/\td$. The prediction by PGH theory  are shown for the respective barrier-friction time as gray and for the well-friction time as colored solid lines. {\bf B, E:} \ac{MFPT} plotted over the well-friction time $\tW/\td$. The prediction by PGH theory is shown for the respective well-friction time in gray and in case of GH theory for the barrier-friction time as colored broken lines. {\bf C, F:} Perturbation parameters $\epsilon_{\rm B}$ for the barrier region and $\epsilon_{\rm W}$ for the well region and energy loss $\Delta E / (k_BT)$, both relevant for stability of the \ac{KSR} theory. The thin black horizontal line denotes the value 1.}
 \label{KSR_results_T}
\end{figure*}

To rationalize the deviations between numerical and analytical predictions, we point out
 that \ac{KSR} themselves state that their theory 
is not to expected to be reliable
 if $\epsilon_{\rm B} \gtrsim 1$ in the barrier region or $\epsilon_{\rm W}\gtrsim 1$ in the well region (note that $\epsilon$ and $\epsilon'$ are used in the original work \cite{Krishnan1992}).
 The perturbation parameters $\epsilon_{\rm B}$, $\epsilon_{\rm W}$ represent a
  measure for the strength of coupling between reaction coordinate
 and  heat bath,
  and are defined as $\epsilon_{\rm B} = \gB/\left(2 m\lambda_{\rm B} (1+ \tB \lambda_{\rm B})^2\right)$ and $\epsilon_{\rm W} = \gW/\left(2 m\lambda_{\rm W} (1+ \tW \lambda_{\rm W})^2\right)$, with $\lambda_{\rm B}$ and $\lambda_{\rm W}$  the Grote/Hynes frequencies which solve of eq.~\eqref{eq:gh_lambda} for the respective 
  memory kernels (well/barrier).
 These conditions are easily violated in case of small inertial time scales $\tm = m/\gamma$,
 as fig.~\ref{KSR_results_T}C shows.
However, a similar perturbation parameter $\epsilon$ is also relevant for the applicability of \ac{PGH} theory,
and PGH note in their paper that PGH theory remains valid even for large $\epsilon$, if at the same time the energy loss per cycle through the well region is large,  $\beta \Delta E>1$ \cite{Pollak1989}. 
More so, the predictions by PGH have been shown to globally agree well
 with numerical results obtained from a homogeneous memory kernel \cite{Kappler2018}. 

The clear deviations between \ac{KSR} theory and the numerical results in fig.~\ref{KSR_results_T}A,
observed for small $\tB/\td$ and large $\tW/\td$ (red and blue lines),
can be rationalized by the simultaneous breakdown of both the conditions on the pair $\epsilon_{\rm B}$, $\epsilon_{\rm W}$, 
and $\beta \Delta E$:
As fig.~\ref{KSR_results_T}C shows, in the regime where deviations between theory and numerical data are observed, $\epsilon_{\rm B} \gtrsim 1$ while $\beta \Delta E \ll 1$. 
In contrast to that fig.~\ref{KSR_results_T}F,
 shows the perturbation parameters and energy loss per cycle for slightly larger inertial
  times $\tm/\td=0.01$.
  Here the conditions $\epsilon_{\rm B}<1$ and $\epsilon_{\rm W}<1$ are met and the predictions agree with the simulation data in fig.~\ref{KSR_results_T}C.

In fig.~\ref{inertialBarrier} we considered the \ac{MFPT} for the cases where $\gB/\gamma \ll 1$ and
$\gW/\gamma \ll 1$, i.e.~the the scenario where the well and barrier-friction magnitudes are very different.
In fig.~\ref{KSR_inertialBarrier}A and C we compare the numerical results for the \ac{MFPT} 
with KSR theory.
As can be observed in fig.~\ref{KSR_inertialBarrier}A, 
 KSR theory (orange dotted line) correctly captures the limit $\gB/\gamma \ll 1$, 
i.e.~the limit of  Markovian low friction on the barrier. 
However, fig.~\ref{KSR_inertialBarrier}C shows that KSR theory does not capture the opposite limit.
For $\gW/\gamma \ll 1$, KSR theory predicts a significant slow-down, 
which is not confirmed by simulation data. 
The breakdown of KSR theory is again understood by considering
 the perturbation parameters $\epsilon_{\rm B}$, $\epsilon_{\rm W}$ and the energy loss $\Delta E/(k_BT)$, which are plotted in fig.~\ref{KSR_inertialBarrier}B and D for the respective data. 
 As previously discussed for the data in fig.~\ref{KSR_results_T}, 
 KSR theory breaks down whenever the energy loss per cycle in the well region is small, $\Delta E/(k_BT) \ll 1$ , while the coupling to the barrier heat bath is strong, $\epsilon_{\rm B} \gtrsim 1$. This is again the case for the data in fig.~\ref{KSR_inertialBarrier}C, as can be seen from the corresponding perturbation parameters in fig.~\ref{KSR_inertialBarrier}D.

In summary, while \ac{KSR} theory does capture the crossover from well-dominated to 
barrier-dominated \acp{MFPT}, the theory captures neither the `Markovian-barrier acceleration' regime, nor the limit $\gW/\gamma \ll 1$.
This can be explained by the assumptions underlying the KSR derivation, which are not fulfilled in these regimes: 
In both regimes, the energy exchange with the well heat bath is weak (small $\beta \Delta E \ll 1$), while simultaneously the coupling to the barrier heat bath is strong ($\epsilon_{\rm B} \gtrsim 1$).
While in the `Markovian-barrier acceleration' regime, the weak energy exchange in the well is due to long memory, in the regime $\gW/\gamma \ll 1$, the weak energy exchange is because of the small well friction.

\begin{figure*}[htb]
  \begin{overpic}[width=0.8\columnwidth]{{/../figs/barrierReset_flo_pub/TM0.001_overGB_log_wGHT_KSR}.pdf}
  \put(0,68){ \huge A}
  \end{overpic}
  \begin{overpic}[width=0.8\columnwidth]{{/../figs/barrierReset_flo_pub/epsDelta_TM0.001_overGB_log}.pdf}
  \put(0,68){\huge B}
  \end{overpic}  
 \begin{overpic}[width=0.8\columnwidth]{{/../figs/barrierReset_flo_pub/TM0.001_overGW_log_KSR}.pdf}
 \put(0,68){ \huge C}
 \end{overpic}
  \begin{overpic}[width=0.8\columnwidth]{{/../figs/barrierReset_flo_pub/epsDelta_TM0.001_overGW_log}.pdf}
  \put(0,68){ \huge D}
  \end{overpic} 
 \caption{\Acl{MFPT}, $\tmfp/\td$, for various single-exponential barrier- and well-friction parameters,
compared with analytic predictions given by \citet{Grote1980} (GH, solid line in A), \citet{Pollak1989} (PGH, evaluated using the heuristic formula \cite{Kappler2019}), transition-state theory (TST, gray broken lines), as well as the predictions by \citet{Krishnan1992} (KSR, dotted lines).
The numerical data is shown for equal barrier and well-friction times, $\tB/\td=\tW/\td=10^{-3}$ and the inertial time scale is fixed in the high-friction regime $\tm/\td=10^{-3}$.
{\bf A:} Results for the limit $\gB/\gamma \ll 1$, i.e.~the limit of Markovian low friction on the barrier.
{\bf C:} Results for the limit $\gW/\gamma \ll 1$, i.e.~the limit of Markovian low friction in the well.
{\bf B, D:} Perturbation parameters $\epsilon_{\rm B}$ for the barrier region and $\epsilon_{\rm W}$ for the well region and energy loss $\Delta E / (k_BT)$, both relevant for stability of the \ac{KSR} theory. The thin black horizontal line denotes the value 1.
} 
 \label{KSR_inertialBarrier}
\end{figure*}

\begin{acronym}[Bash]
 \acro{FPT}{first-passage time}
 \acro{GLE}{generalized Langevin equation}
 \acro{GH}{Grote and Hynes}
 \acro{GHT}{Grote-Hynes theory}
 \acro{KSR}{Krishnan, Singh and Robinson}
 \acro{LE}{Langevin equation}
 \acro{MD}{molecular dynamics}
 \acro{MFPT}{mean first-passage time}
 \acro{MFP}{mean first-passage}
 \acro{MM}{Mel'nikov and Meshkov}
 \acro{MSD}{mean squared displacement}
 \acro{PGH}{Pollak, Grabert and Hanggi}
 \acro{TP}{transition path}
\end{acronym}

\clearpage

%


\begin{thebibliography}{56}%
\makeatletter
\providecommand \@ifxundefined [1]{%
 \@ifx{#1\undefined}
}%
\providecommand \@ifnum [1]{%
 \ifnum #1\expandafter \@firstoftwo
 \else \expandafter \@secondoftwo
 \fi
}%
\providecommand \@ifx [1]{%
 \ifx #1\expandafter \@firstoftwo
 \else \expandafter \@secondoftwo
 \fi
}%
\providecommand \natexlab [1]{#1}%
\providecommand \enquote  [1]{``#1''}%
\providecommand \bibnamefont  [1]{#1}%
\providecommand \bibfnamefont [1]{#1}%
\providecommand \citenamefont [1]{#1}%
\providecommand \href@noop [0]{\@secondoftwo}%
\providecommand \href [0]{\begingroup \@sanitize@url \@href}%
\providecommand \@href[1]{\@@startlink{#1}\@@href}%
\providecommand \@@href[1]{\endgroup#1\@@endlink}%
\providecommand \@sanitize@url [0]{\catcode `\\12\catcode `\$12\catcode
  `\&12\catcode `\#12\catcode `\^12\catcode `\_12\catcode `\%12\relax}%
\providecommand \@@startlink[1]{}%
\providecommand \@@endlink[0]{}%
\providecommand \url  [0]{\begingroup\@sanitize@url \@url }%
\providecommand \@url [1]{\endgroup\@href {#1}{\urlprefix }}%
\providecommand \urlprefix  [0]{URL }%
\providecommand \Eprint [0]{\href }%
\providecommand \doibase [0]{https://doi.org/}%
\providecommand \selectlanguage [0]{\@gobble}%
\providecommand \bibinfo  [0]{\@secondoftwo}%
\providecommand \bibfield  [0]{\@secondoftwo}%
\providecommand \translation [1]{[#1]}%
\providecommand \BibitemOpen [0]{}%
\providecommand \bibitemStop [0]{}%
\providecommand \bibitemNoStop [0]{.\EOS\space}%
\providecommand \EOS [0]{\spacefactor3000\relax}%
\providecommand \BibitemShut  [1]{\csname bibitem#1\endcsname}%
\let\auto@bib@innerbib\@empty
\bibitem [{\citenamefont {Zwanzig}(1961)}]{Zwanzig1961}%
  \BibitemOpen
  \bibfield  {author} {\bibinfo {author} {\bibfnamefont {R.}~\bibnamefont
  {Zwanzig}},\ }\bibfield  {title} {\bibinfo {title} {{Memory effects in
  irreversible thermodynamics}},\ }\href
  {https://doi.org/10.1103/PhysRev.124.983} {\bibfield  {journal} {\bibinfo
  {journal} {Phys. Rev.}\ }\textbf {\bibinfo {volume} {124}},\ \bibinfo {pages}
  {983} (\bibinfo {year} {1961})}\BibitemShut {NoStop}%
\bibitem [{\citenamefont {Mori}(1965)}]{Mori1965a}%
  \BibitemOpen
  \bibfield  {author} {\bibinfo {author} {\bibfnamefont {H.}~\bibnamefont
  {Mori}},\ }\bibfield  {title} {\bibinfo {title} {{Transport, collective
  motion, and Brownian motion}},\ }\href {https://doi.org/10.1143/PTP.33.423}
  {\bibfield  {journal} {\bibinfo  {journal} {Prog. Theor. Phys.}\ }\textbf
  {\bibinfo {volume} {33}},\ \bibinfo {pages} {423} (\bibinfo {year}
  {1965})}\BibitemShut {NoStop}%
\bibitem [{\citenamefont {Zwanzig}(1973)}]{Zwanzig1973}%
  \BibitemOpen
  \bibfield  {author} {\bibinfo {author} {\bibfnamefont {R.}~\bibnamefont
  {Zwanzig}},\ }\bibfield  {title} {\bibinfo {title} {{Nonlinear generalized
  Langevin equations}},\ }\href {https://doi.org/10.1007/BF01008729} {\bibfield
   {journal} {\bibinfo  {journal} {J. Stat. Phys.}\ }\textbf {\bibinfo {volume}
  {9}},\ \bibinfo {pages} {215} (\bibinfo {year} {1973})}\BibitemShut {NoStop}%
\bibitem [{\citenamefont {Zwanzig}(2001)}]{Zwanzig2001}%
  \BibitemOpen
  \bibfield  {author} {\bibinfo {author} {\bibfnamefont {R.}~\bibnamefont
  {Zwanzig}},\ }\href@noop {} {\emph {\bibinfo {title} {{Nonequilibrium
  Statistical Mechanics}}}}\ (\bibinfo  {publisher} {Oxford University Press},\
  \bibinfo {year} {2001})\BibitemShut {NoStop}%
\bibitem [{\citenamefont {Berezhkovskii}\ and\ \citenamefont
  {Szabo}(2005)}]{Berezhkovskii2005}%
  \BibitemOpen
  \bibfield  {author} {\bibinfo {author} {\bibfnamefont {A.}~\bibnamefont
  {Berezhkovskii}}\ and\ \bibinfo {author} {\bibfnamefont {A.}~\bibnamefont
  {Szabo}},\ }\bibfield  {title} {\bibinfo {title} {{One-dimensional reaction
  coordinates for diffusive activated rate processes in many dimensions}},\
  }\href {https://doi.org/10.1063/1.1818091} {\bibfield  {journal} {\bibinfo
  {journal} {J. Chem. Phys.}\ }\textbf {\bibinfo {volume} {122}},\ \bibinfo
  {pages} {14503} (\bibinfo {year} {2005})}\BibitemShut {NoStop}%
\bibitem [{\citenamefont {Lange}\ and\ \citenamefont
  {Grubm{\"{u}}ller}(2006)}]{Lange2006}%
  \BibitemOpen
  \bibfield  {author} {\bibinfo {author} {\bibfnamefont {O.~F.}\ \bibnamefont
  {Lange}}\ and\ \bibinfo {author} {\bibfnamefont {H.}~\bibnamefont
  {Grubm{\"{u}}ller}},\ }\bibfield  {title} {\bibinfo {title} {{Collective
  Langevin dynamics of conformational motions in proteins}},\ }\href
  {https://doi.org/10.1063/1.2199530} {\bibfield  {journal} {\bibinfo
  {journal} {J. Chem. Phys.}\ }\textbf {\bibinfo {volume} {124}},\ \bibinfo
  {pages} {214903} (\bibinfo {year} {2006})}\BibitemShut {NoStop}%
\bibitem [{\citenamefont {Daldrop}\ \emph {et~al.}(2018)\citenamefont
  {Daldrop}, \citenamefont {Kappler}, \citenamefont {Br{\"{u}}nig},\ and\
  \citenamefont {Netz}}]{Daldrop2018}%
  \BibitemOpen
  \bibfield  {author} {\bibinfo {author} {\bibfnamefont {J.~O.}\ \bibnamefont
  {Daldrop}}, \bibinfo {author} {\bibfnamefont {J.}~\bibnamefont {Kappler}},
  \bibinfo {author} {\bibfnamefont {F.~N.}\ \bibnamefont {Br{\"{u}}nig}},\ and\
  \bibinfo {author} {\bibfnamefont {R.~R.}\ \bibnamefont {Netz}},\ }\bibfield
  {title} {\bibinfo {title} {{Butane dihedral angle dynamics in water is
  dominated by internal friction}},\ }\href
  {https://doi.org/10.1073/pnas.1722327115} {\bibfield  {journal} {\bibinfo
  {journal} {Proc. Natl. Acad. Sci.}\ }\textbf {\bibinfo {volume} {115}},\
  \bibinfo {pages} {5169} (\bibinfo {year} {2018})}\BibitemShut {NoStop}%
\bibitem [{\citenamefont {Kappler}\ \emph
  {et~al.}(2019{\natexlab{a}})\citenamefont {Kappler}, \citenamefont
  {No{\'{e}}},\ and\ \citenamefont {Netz}}]{Kappler2019b}%
  \BibitemOpen
  \bibfield  {author} {\bibinfo {author} {\bibfnamefont {J.}~\bibnamefont
  {Kappler}}, \bibinfo {author} {\bibfnamefont {F.}~\bibnamefont {No{\'{e}}}},\
  and\ \bibinfo {author} {\bibfnamefont {R.~R.}\ \bibnamefont {Netz}},\
  }\bibfield  {title} {\bibinfo {title} {{Cyclization and Relaxation Dynamics
  of Finite-Length Collapsed Self-Avoiding Polymers}},\ }\href
  {https://doi.org/10.1103/PhysRevLett.122.067801} {\bibfield  {journal}
  {\bibinfo  {journal} {Phys. Rev. Lett.}\ }\textbf {\bibinfo {volume} {122}},\
  \bibinfo {pages} {067801} (\bibinfo {year} {2019}{\natexlab{a}})}\BibitemShut
  {NoStop}%
\bibitem [{\citenamefont {Satija}\ and\ \citenamefont
  {Makarov}(2019)}]{Satija2019}%
  \BibitemOpen
  \bibfield  {author} {\bibinfo {author} {\bibfnamefont {R.}~\bibnamefont
  {Satija}}\ and\ \bibinfo {author} {\bibfnamefont {D.~E.}\ \bibnamefont
  {Makarov}},\ }\bibfield  {title} {\bibinfo {title} {{Generalized Langevin
  Equation as a Model for Barrier Crossing Dynamics in Biomolecular Folding}},\
  }\href {https://doi.org/10.1021/acs.jpcb.8b11137} {\bibfield  {journal}
  {\bibinfo  {journal} {J. Phys. Chem. B}\ }\textbf {\bibinfo {volume} {123}},\
  \bibinfo {pages} {802} (\bibinfo {year} {2019})}\BibitemShut {NoStop}%
\bibitem [{\citenamefont {Lickert}\ and\ \citenamefont
  {Stock}(2020)}]{Lickert2020}%
  \BibitemOpen
  \bibfield  {author} {\bibinfo {author} {\bibfnamefont {B.}~\bibnamefont
  {Lickert}}\ and\ \bibinfo {author} {\bibfnamefont {G.}~\bibnamefont
  {Stock}},\ }\bibfield  {title} {\bibinfo {title} {{Modeling non-Markovian
  data using Markov state and Langevin models}},\ }\href
  {https://doi.org/10.1063/5.0031979} {\bibfield  {journal} {\bibinfo
  {journal} {J. Chem. Phys.}\ }\textbf {\bibinfo {volume} {153}},\ \bibinfo
  {pages} {244112} (\bibinfo {year} {2020})}\BibitemShut {NoStop}%
\bibitem [{\citenamefont {Kramers}(1940)}]{Kramers1940}%
  \BibitemOpen
  \bibfield  {author} {\bibinfo {author} {\bibfnamefont {H.}~\bibnamefont
  {Kramers}},\ }\bibfield  {title} {\bibinfo {title} {{Brownian motion in a
  field of force and the diffusion model of chemical reactions}},\ }\href
  {https://doi.org/10.1016/S0031-8914(40)90098-2} {\bibfield  {journal}
  {\bibinfo  {journal} {Physica}\ }\textbf {\bibinfo {volume} {7}},\ \bibinfo
  {pages} {284} (\bibinfo {year} {1940})}\BibitemShut {NoStop}%
\bibitem [{\citenamefont {Chandler}(1978)}]{Chandler1978}%
  \BibitemOpen
  \bibfield  {author} {\bibinfo {author} {\bibfnamefont {D.}~\bibnamefont
  {Chandler}},\ }\bibfield  {title} {\bibinfo {title} {{Statistical mechanics
  of isomerization dynamics in liquids and the transition state
  approximation}},\ }\href {https://doi.org/10.1063/1.436049} {\bibfield
  {journal} {\bibinfo  {journal} {J. Chem. Phys.}\ }\textbf {\bibinfo {volume}
  {68}},\ \bibinfo {pages} {2959} (\bibinfo {year} {1978})}\BibitemShut
  {NoStop}%
\bibitem [{\citenamefont {Grote}\ and\ \citenamefont
  {Hynes}(1980)}]{Grote1980}%
  \BibitemOpen
  \bibfield  {author} {\bibinfo {author} {\bibfnamefont {R.~F.}\ \bibnamefont
  {Grote}}\ and\ \bibinfo {author} {\bibfnamefont {J.~T.}\ \bibnamefont
  {Hynes}},\ }\bibfield  {title} {\bibinfo {title} {{The stable states picture
  of chemical reactions. II. Rate constants for condensed and gas phase
  reaction models}},\ }\href {https://doi.org/10.1063/1.440485} {\bibfield
  {journal} {\bibinfo  {journal} {J. Chem. Phys.}\ }\textbf {\bibinfo {volume}
  {73}},\ \bibinfo {pages} {2715} (\bibinfo {year} {1980})}\BibitemShut
  {NoStop}%
\bibitem [{\citenamefont {Chandler}(1986)}]{Chandler1986}%
  \BibitemOpen
  \bibfield  {author} {\bibinfo {author} {\bibfnamefont {D.}~\bibnamefont
  {Chandler}},\ }\bibfield  {title} {\bibinfo {title} {{Roles of classical
  dynamics and quantum dynamics on activated processes occurring in liquids}},\
  }\href {https://doi.org/10.1007/BF01010840} {\bibfield  {journal} {\bibinfo
  {journal} {J. Stat. Phys.}\ }\textbf {\bibinfo {volume} {42}},\ \bibinfo
  {pages} {49} (\bibinfo {year} {1986})}\BibitemShut {NoStop}%
\bibitem [{\citenamefont {Mel'nikov}\ and\ \citenamefont
  {Meshkov}(1986)}]{Melnikov1986}%
  \BibitemOpen
  \bibfield  {author} {\bibinfo {author} {\bibfnamefont {V.~I.}\ \bibnamefont
  {Mel'nikov}}\ and\ \bibinfo {author} {\bibfnamefont {S.~V.}\ \bibnamefont
  {Meshkov}},\ }\bibfield  {title} {\bibinfo {title} {{Theory of activated rate
  processes: Exact solution of the Kramers problem}},\ }\href
  {https://doi.org/10.1063/1.451844} {\bibfield  {journal} {\bibinfo  {journal}
  {J. Chem. Phys.}\ }\textbf {\bibinfo {volume} {85}},\ \bibinfo {pages} {1018}
  (\bibinfo {year} {1986})}\BibitemShut {NoStop}%
\bibitem [{\citenamefont {H{\"{a}}nggi}\ \emph {et~al.}(1990)\citenamefont
  {H{\"{a}}nggi}, \citenamefont {Talkner},\ and\ \citenamefont
  {Borkovec}}]{Hanggi1990}%
  \BibitemOpen
  \bibfield  {author} {\bibinfo {author} {\bibfnamefont {P.}~\bibnamefont
  {H{\"{a}}nggi}}, \bibinfo {author} {\bibfnamefont {P.}~\bibnamefont
  {Talkner}},\ and\ \bibinfo {author} {\bibfnamefont {M.}~\bibnamefont
  {Borkovec}},\ }\bibfield  {title} {\bibinfo {title} {{Reaction-rate theory:
  Fifty years after Kramers}},\ }\href
  {https://doi.org/10.1103/RevModPhys.62.251} {\bibfield  {journal} {\bibinfo
  {journal} {Rev. Mod. Phys.}\ }\textbf {\bibinfo {volume} {62}},\ \bibinfo
  {pages} {251} (\bibinfo {year} {1990})}\BibitemShut {NoStop}%
\bibitem [{\citenamefont {Mel'nikov}(1991)}]{Melnikov1991}%
  \BibitemOpen
  \bibfield  {author} {\bibinfo {author} {\bibfnamefont {V.~I.}\ \bibnamefont
  {Mel'nikov}},\ }\bibfield  {title} {\bibinfo {title} {{The Kramers problem:
  Fifty years of development}},\ }\href
  {https://doi.org/10.1016/0370-1573(91)90108-X} {\bibfield  {journal}
  {\bibinfo  {journal} {Phys. Rep.}\ }\textbf {\bibinfo {volume} {209}},\
  \bibinfo {pages} {1} (\bibinfo {year} {1991})}\BibitemShut {NoStop}%
\bibitem [{\citenamefont {Best}\ and\ \citenamefont {Hummer}(2006)}]{Best2006}%
  \BibitemOpen
  \bibfield  {author} {\bibinfo {author} {\bibfnamefont {R.}~\bibnamefont
  {Best}}\ and\ \bibinfo {author} {\bibfnamefont {G.}~\bibnamefont {Hummer}},\
  }\bibfield  {title} {\bibinfo {title} {{Diffusive {Model} of {Protein}
  {Folding} {Dynamics} with {Kramers} {Turnover} in {Rate}}},\ }\bibfield
  {journal} {\bibinfo  {journal} {Phys. Rev. Lett.}\ }\textbf {\bibinfo
  {volume} {96}},\ \href {https://doi.org/10.1103/PhysRevLett.96.228104}
  {10.1103/PhysRevLett.96.228104} (\bibinfo {year} {2006})\BibitemShut
  {NoStop}%
\bibitem [{\citenamefont {Wilemski}\ and\ \citenamefont
  {Fixman}(1974)}]{Wilemski1974}%
  \BibitemOpen
  \bibfield  {author} {\bibinfo {author} {\bibfnamefont {G.}~\bibnamefont
  {Wilemski}}\ and\ \bibinfo {author} {\bibfnamefont {M.}~\bibnamefont
  {Fixman}},\ }\bibfield  {title} {\bibinfo {title} {{Diffusion-controlled
  intrachain reactions of polymers. II Results for a pair of terminal reactive
  groups}},\ }\href {https://doi.org/10.1063/1.1681163} {\bibfield  {journal}
  {\bibinfo  {journal} {J. Chem. Phys.}\ }\textbf {\bibinfo {volume} {60}},\
  \bibinfo {pages} {878} (\bibinfo {year} {1974})}\BibitemShut {NoStop}%
\bibitem [{\citenamefont {Szabo}\ \emph {et~al.}(1980)\citenamefont {Szabo},
  \citenamefont {Schulten},\ and\ \citenamefont {Schulten}}]{Szabo1980}%
  \BibitemOpen
  \bibfield  {author} {\bibinfo {author} {\bibfnamefont {A.}~\bibnamefont
  {Szabo}}, \bibinfo {author} {\bibfnamefont {K.}~\bibnamefont {Schulten}},\
  and\ \bibinfo {author} {\bibfnamefont {Z.}~\bibnamefont {Schulten}},\
  }\bibfield  {title} {\bibinfo {title} {{First passage time approach to
  diffusion controlled reactions}},\ }\href {https://doi.org/10.1063/1.439715}
  {\bibfield  {journal} {\bibinfo  {journal} {J. Chem. Phys.}\ }\textbf
  {\bibinfo {volume} {72}},\ \bibinfo {pages} {4350} (\bibinfo {year}
  {1980})}\BibitemShut {NoStop}%
\bibitem [{\citenamefont {Kappler}\ \emph
  {et~al.}(2019{\natexlab{b}})\citenamefont {Kappler}, \citenamefont
  {Hinrichsen},\ and\ \citenamefont {Netz}}]{Kappler2019}%
  \BibitemOpen
  \bibfield  {author} {\bibinfo {author} {\bibfnamefont {J.}~\bibnamefont
  {Kappler}}, \bibinfo {author} {\bibfnamefont {V.~B.}\ \bibnamefont
  {Hinrichsen}},\ and\ \bibinfo {author} {\bibfnamefont {R.~R.}\ \bibnamefont
  {Netz}},\ }\bibfield  {title} {\bibinfo {title} {{Non-Markovian barrier
  crossing with two-time-scale memory is dominated by the faster memory
  component}},\ }\href@noop {} {\bibfield  {journal} {\bibinfo  {journal} {Eur.
  Phys. J. E}\ }\textbf {\bibinfo {volume} {42}},\ \bibinfo {pages} {119}
  (\bibinfo {year} {2019}{\natexlab{b}})}\BibitemShut {NoStop}%
\bibitem [{\citenamefont {Ayaz}\ \emph {et~al.}(2021)\citenamefont {Ayaz},
  \citenamefont {Tepper}, \citenamefont {Br{\"{u}}nig}, \citenamefont
  {Kappler}, \citenamefont {Daldrop},\ and\ \citenamefont {Netz}}]{Ayaz2021}%
  \BibitemOpen
  \bibfield  {author} {\bibinfo {author} {\bibfnamefont {C.}~\bibnamefont
  {Ayaz}}, \bibinfo {author} {\bibfnamefont {L.}~\bibnamefont {Tepper}},
  \bibinfo {author} {\bibfnamefont {F.~N.}\ \bibnamefont {Br{\"{u}}nig}},
  \bibinfo {author} {\bibfnamefont {J.}~\bibnamefont {Kappler}}, \bibinfo
  {author} {\bibfnamefont {J.~O.}\ \bibnamefont {Daldrop}},\ and\ \bibinfo
  {author} {\bibfnamefont {R.~R.}\ \bibnamefont {Netz}},\ }\bibfield  {title}
  {\bibinfo {title} {{Non-Markovian modeling of protein folding}},\ }\href
  {https://doi.org/10.1073/pnas.2023856118} {\bibfield  {journal} {\bibinfo
  {journal} {Proc. Natl. Acad. Sci. U. S. A.}\ }\textbf {\bibinfo {volume}
  {118}},\ \bibinfo {pages} {20} (\bibinfo {year} {2021})}\BibitemShut
  {NoStop}%
\bibitem [{\citenamefont {Straub}\ \emph {et~al.}(1987)\citenamefont {Straub},
  \citenamefont {Borkovec},\ and\ \citenamefont {Berne}}]{Straub1987}%
  \BibitemOpen
  \bibfield  {author} {\bibinfo {author} {\bibfnamefont {J.~E.}\ \bibnamefont
  {Straub}}, \bibinfo {author} {\bibfnamefont {M.}~\bibnamefont {Borkovec}},\
  and\ \bibinfo {author} {\bibfnamefont {B.~J.}\ \bibnamefont {Berne}},\
  }\bibfield  {title} {\bibinfo {title} {{Calculation of dynamic friction on
  intramolecular degrees of freedom}},\ }\href
  {https://doi.org/10.1021/j100303a019} {\bibfield  {journal} {\bibinfo
  {journal} {J. Phys. Chem.}\ }\textbf {\bibinfo {volume} {91}},\ \bibinfo
  {pages} {4995} (\bibinfo {year} {1987})}\BibitemShut {NoStop}%
\bibitem [{\citenamefont {Ciccotti}\ \emph {et~al.}(1990)\citenamefont
  {Ciccotti}, \citenamefont {Ferrario}, \citenamefont {Hynes},\ and\
  \citenamefont {Kapral}}]{Ciccotti1990}%
  \BibitemOpen
  \bibfield  {author} {\bibinfo {author} {\bibfnamefont {G.}~\bibnamefont
  {Ciccotti}}, \bibinfo {author} {\bibfnamefont {M.}~\bibnamefont {Ferrario}},
  \bibinfo {author} {\bibfnamefont {J.~T.}\ \bibnamefont {Hynes}},\ and\
  \bibinfo {author} {\bibfnamefont {R.}~\bibnamefont {Kapral}},\ }\bibfield
  {title} {\bibinfo {title} {{Dynamics of ion pair interconversion in a polar
  solvent}},\ }\href {https://doi.org/10.1063/1.459437} {\bibfield  {journal}
  {\bibinfo  {journal} {J. Chem. Phys.}\ }\textbf {\bibinfo {volume} {93}},\
  \bibinfo {pages} {7137} (\bibinfo {year} {1990})}\BibitemShut {NoStop}%
\bibitem [{\citenamefont {Benjamin}\ \emph {et~al.}(1991)\citenamefont
  {Benjamin}, \citenamefont {Lee}, \citenamefont {Li}, \citenamefont {Liu},\
  and\ \citenamefont {Wilson}}]{Benjamin1991}%
  \BibitemOpen
  \bibfield  {author} {\bibinfo {author} {\bibfnamefont {I.}~\bibnamefont
  {Benjamin}}, \bibinfo {author} {\bibfnamefont {L.~L.}\ \bibnamefont {Lee}},
  \bibinfo {author} {\bibfnamefont {Y.~S.}\ \bibnamefont {Li}}, \bibinfo
  {author} {\bibfnamefont {A.}~\bibnamefont {Liu}},\ and\ \bibinfo {author}
  {\bibfnamefont {K.~R.}\ \bibnamefont {Wilson}},\ }\bibfield  {title}
  {\bibinfo {title} {{Generalized Langevin model for molecular dynamics of an
  activated reaction in solution}},\ }\href
  {https://doi.org/10.1016/0301-0104(91)80029-H} {\bibfield  {journal}
  {\bibinfo  {journal} {Chem. Phys.}\ }\textbf {\bibinfo {volume} {152}},\
  \bibinfo {pages} {1} (\bibinfo {year} {1991})}\BibitemShut {NoStop}%
\bibitem [{\citenamefont {Rey}\ \emph {et~al.}(1992)\citenamefont {Rey},
  \citenamefont {Gu{\`{a}}rdia},\ and\ \citenamefont {Padr{\'{o}}}}]{Rey1992a}%
  \BibitemOpen
  \bibfield  {author} {\bibinfo {author} {\bibfnamefont {R.}~\bibnamefont
  {Rey}}, \bibinfo {author} {\bibfnamefont {E.}~\bibnamefont {Gu{\`{a}}rdia}},\
  and\ \bibinfo {author} {\bibfnamefont {J.~A.}\ \bibnamefont {Padr{\'{o}}}},\
  }\bibfield  {title} {\bibinfo {title} {{Generalized Langevin dynamics
  simulation of activated processes in solution: Ion pair interconversion in
  water}},\ }\href {https://doi.org/10.1063/1.463398} {\bibfield  {journal}
  {\bibinfo  {journal} {J. Chem. Phys.}\ }\textbf {\bibinfo {volume} {97}},\
  \bibinfo {pages} {8276} (\bibinfo {year} {1992})}\BibitemShut {NoStop}%
\bibitem [{\citenamefont {Annapureddy}\ and\ \citenamefont
  {Dang}(2014)}]{Annapureddy2014}%
  \BibitemOpen
  \bibfield  {author} {\bibinfo {author} {\bibfnamefont {H.~V.~R.}\
  \bibnamefont {Annapureddy}}\ and\ \bibinfo {author} {\bibfnamefont {L.~X.}\
  \bibnamefont {Dang}},\ }\bibfield  {title} {\bibinfo {title} {{Understanding
  the Rates and Molecular Mechanism of Water-Exchange around Aqueous Ions Using
  Molecular Simulations}},\ }\href@noop {} {\bibfield  {journal} {\bibinfo
  {journal} {J. Phys. Chem. B}\ }\textbf {\bibinfo {volume} {118}},\ \bibinfo
  {pages} {8917} (\bibinfo {year} {2014})}\BibitemShut {NoStop}%
\bibitem [{\citenamefont {Meyer}\ \emph {et~al.}(2021)\citenamefont {Meyer},
  \citenamefont {Wolf}, \citenamefont {Stock},\ and\ \citenamefont
  {Schilling}}]{Meyer2021}%
  \BibitemOpen
  \bibfield  {author} {\bibinfo {author} {\bibfnamefont {H.}~\bibnamefont
  {Meyer}}, \bibinfo {author} {\bibfnamefont {S.}~\bibnamefont {Wolf}},
  \bibinfo {author} {\bibfnamefont {G.}~\bibnamefont {Stock}},\ and\ \bibinfo
  {author} {\bibfnamefont {T.}~\bibnamefont {Schilling}},\ }\bibfield  {title}
  {\bibinfo {title} {{A Numerical Procedure to Evaluate Memory Effects in
  Non-Equilibrium Coarse-Grained Models}},\ }\href
  {https://doi.org/10.1002/adts.202000197} {\bibfield  {journal} {\bibinfo
  {journal} {Adv. Theory Simulations}\ }\textbf {\bibinfo {volume} {4}},\
  \bibinfo {pages} {2000197} (\bibinfo {year} {2021})}\BibitemShut {NoStop}%
\bibitem [{\citenamefont {Zwanzig}(1960)}]{Zwanzig1960}%
  \BibitemOpen
  \bibfield  {author} {\bibinfo {author} {\bibfnamefont {R.}~\bibnamefont
  {Zwanzig}},\ }\bibfield  {title} {\bibinfo {title} {{Ensemble Method in the
  Theory of Irreversibility}},\ }\href {https://doi.org/10.1063/1.1731409}
  {\bibfield  {journal} {\bibinfo  {journal} {J. Chem. Phys.}\ }\textbf
  {\bibinfo {volume} {33}},\ \bibinfo {pages} {1338} (\bibinfo {year}
  {1960})}\BibitemShut {NoStop}%
\bibitem [{\citenamefont {Daldrop}\ \emph {et~al.}(2017)\citenamefont
  {Daldrop}, \citenamefont {Kowalik},\ and\ \citenamefont
  {Netz}}]{Daldrop2017a}%
  \BibitemOpen
  \bibfield  {author} {\bibinfo {author} {\bibfnamefont {J.~O.}\ \bibnamefont
  {Daldrop}}, \bibinfo {author} {\bibfnamefont {B.~G.}\ \bibnamefont
  {Kowalik}},\ and\ \bibinfo {author} {\bibfnamefont {R.~R.}\ \bibnamefont
  {Netz}},\ }\bibfield  {title} {\bibinfo {title} {{External Potential Modifies
  Friction of Molecular Solutes in Water}},\ }\href
  {https://doi.org/10.1103/PhysRevX.7.041065} {\bibfield  {journal} {\bibinfo
  {journal} {Phys. Rev. X}\ }\textbf {\bibinfo {volume} {7}},\ \bibinfo {pages}
  {41065} (\bibinfo {year} {2017})}\BibitemShut {NoStop}%
\bibitem [{\citenamefont {M{\"{u}}ller}\ \emph {et~al.}(2020)\citenamefont
  {M{\"{u}}ller}, \citenamefont {Berner}, \citenamefont {Bechinger},\ and\
  \citenamefont {Kr{\"{u}}ger}}]{Muller2020}%
  \BibitemOpen
  \bibfield  {author} {\bibinfo {author} {\bibfnamefont {B.}~\bibnamefont
  {M{\"{u}}ller}}, \bibinfo {author} {\bibfnamefont {J.}~\bibnamefont
  {Berner}}, \bibinfo {author} {\bibfnamefont {C.}~\bibnamefont {Bechinger}},\
  and\ \bibinfo {author} {\bibfnamefont {M.}~\bibnamefont {Kr{\"{u}}ger}},\
  }\bibfield  {title} {\bibinfo {title} {{Properties of a nonlinear bath:
  experiments, theory, and a stochastic Prandtl-Tomlinson model}},\ }\href
  {https://doi.org/10.1088/1367-2630/ab6a39} {\bibfield  {journal} {\bibinfo
  {journal} {New J. Phys.}\ }\textbf {\bibinfo {volume} {22}},\ \bibinfo
  {pages} {023014} (\bibinfo {year} {2020})}\BibitemShut {NoStop}%
\bibitem [{\citenamefont {Pollak}\ \emph {et~al.}(1989)\citenamefont {Pollak},
  \citenamefont {Grabert},\ and\ \citenamefont {H\"{a}nggi}}]{Pollak1989}%
  \BibitemOpen
  \bibfield  {author} {\bibinfo {author} {\bibfnamefont {E.}~\bibnamefont
  {Pollak}}, \bibinfo {author} {\bibfnamefont {H.}~\bibnamefont {Grabert}},\
  and\ \bibinfo {author} {\bibfnamefont {P.}~\bibnamefont {H\"{a}nggi}},\
  }\bibfield  {title} {\bibinfo {title} {{Theory of activated rate processes
  for arbitrary frequency dependent friction: solution of the turnover
  problem}},\ }\href {https://doi.org/10.1063/1.456837} {\bibfield  {journal}
  {\bibinfo  {journal} {J. Chem. Phys.}\ }\textbf {\bibinfo {volume} {91}},\
  \bibinfo {pages} {4073} (\bibinfo {year} {1989})}\BibitemShut {NoStop}%
\bibitem [{\citenamefont {Kappler}\ \emph {et~al.}(2018)\citenamefont
  {Kappler}, \citenamefont {Daldrop}, \citenamefont {Br{\"{u}}nig},
  \citenamefont {Boehle},\ and\ \citenamefont {Netz}}]{Kappler2018}%
  \BibitemOpen
  \bibfield  {author} {\bibinfo {author} {\bibfnamefont {J.}~\bibnamefont
  {Kappler}}, \bibinfo {author} {\bibfnamefont {J.~O.}\ \bibnamefont
  {Daldrop}}, \bibinfo {author} {\bibfnamefont {F.~N.}\ \bibnamefont
  {Br{\"{u}}nig}}, \bibinfo {author} {\bibfnamefont {M.~D.}\ \bibnamefont
  {Boehle}},\ and\ \bibinfo {author} {\bibfnamefont {R.~R.}\ \bibnamefont
  {Netz}},\ }\bibfield  {title} {\bibinfo {title} {{Memory-induced acceleration
  and slowdown of barrier crossing}},\ }\href
  {https://doi.org/10.1063/1.4998239} {\bibfield  {journal} {\bibinfo
  {journal} {J. Chem. Phys.}\ }\textbf {\bibinfo {volume} {148}},\ \bibinfo
  {pages} {014903} (\bibinfo {year} {2018})}\BibitemShut {NoStop}%
\bibitem [{\citenamefont {Talkner}\ and\ \citenamefont
  {Braun}(1988)}]{Talkner1988}%
  \BibitemOpen
  \bibfield  {author} {\bibinfo {author} {\bibfnamefont {P.}~\bibnamefont
  {Talkner}}\ and\ \bibinfo {author} {\bibfnamefont {H.-B.}\ \bibnamefont
  {Braun}},\ }\bibfield  {title} {\bibinfo {title} {{Transition rates of a
  non-{Markovian} {Brownian} particle in a double well potential}},\ }\href
  {https://doi.org/10.1063/1.454318} {\bibfield  {journal} {\bibinfo  {journal}
  {J. Chem. Phys.}\ }\textbf {\bibinfo {volume} {88}},\ \bibinfo {pages} {7537}
  (\bibinfo {year} {1988})}\BibitemShut {NoStop}%
\bibitem [{\citenamefont {Straub}\ \emph {et~al.}(1986)\citenamefont {Straub},
  \citenamefont {Borkovec},\ and\ \citenamefont {Berne}}]{Straub1986}%
  \BibitemOpen
  \bibfield  {author} {\bibinfo {author} {\bibfnamefont {J.~E.}\ \bibnamefont
  {Straub}}, \bibinfo {author} {\bibfnamefont {M.}~\bibnamefont {Borkovec}},\
  and\ \bibinfo {author} {\bibfnamefont {B.~J.}\ \bibnamefont {Berne}},\
  }\bibfield  {title} {\bibinfo {title} {{Non-Markovian activated rate
  processes: Comparison of current theories with numerical simulation data}},\
  }\href {https://doi.org/10.1063/1.450425} {\bibfield  {journal} {\bibinfo
  {journal} {J. Chem. Phys.}\ }\textbf {\bibinfo {volume} {84}},\ \bibinfo
  {pages} {1788} (\bibinfo {year} {1986})}\BibitemShut {NoStop}%
\bibitem [{\citenamefont {Tucker}\ \emph {et~al.}(1991)\citenamefont {Tucker},
  \citenamefont {Tuckerman}, \citenamefont {Berne},\ and\ \citenamefont
  {Pollak}}]{Tucker1991}%
  \BibitemOpen
  \bibfield  {author} {\bibinfo {author} {\bibfnamefont {S.~C.}\ \bibnamefont
  {Tucker}}, \bibinfo {author} {\bibfnamefont {M.~E.}\ \bibnamefont
  {Tuckerman}}, \bibinfo {author} {\bibfnamefont {B.~J.}\ \bibnamefont
  {Berne}},\ and\ \bibinfo {author} {\bibfnamefont {E.}~\bibnamefont
  {Pollak}},\ }\bibfield  {title} {\bibinfo {title} {{Comparison of rate
  theories for generalized {Langevin} dynamics}},\ }\href
  {https://doi.org/10.1063/1.461603} {\bibfield  {journal} {\bibinfo  {journal}
  {J. Chem. Phys.}\ }\textbf {\bibinfo {volume} {95}},\ \bibinfo {pages} {5809}
  (\bibinfo {year} {1991})}\BibitemShut {NoStop}%
\bibitem [{\citenamefont {Ianconescu}\ and\ \citenamefont
  {Pollak}(2015)}]{Ianconescu2015}%
  \BibitemOpen
  \bibfield  {author} {\bibinfo {author} {\bibfnamefont {R.}~\bibnamefont
  {Ianconescu}}\ and\ \bibinfo {author} {\bibfnamefont {E.}~\bibnamefont
  {Pollak}},\ }\bibfield  {title} {\bibinfo {title} {{A study of
  {Kramers}{\textquoteright} turnover theory in the presence of exponential
  memory friction}},\ }\href {https://doi.org/10.1063/1.4929709} {\bibfield
  {journal} {\bibinfo  {journal} {J. Chem. Phys.}\ }\textbf {\bibinfo {volume}
  {143}},\ \bibinfo {pages} {104104} (\bibinfo {year} {2015})}\BibitemShut
  {NoStop}%
\bibitem [{\citenamefont {Lavacchi}\ \emph {et~al.}(2020)\citenamefont
  {Lavacchi}, \citenamefont {Kappler},\ and\ \citenamefont
  {Netz}}]{Lavacchi2020}%
  \BibitemOpen
  \bibfield  {author} {\bibinfo {author} {\bibfnamefont {L.}~\bibnamefont
  {Lavacchi}}, \bibinfo {author} {\bibfnamefont {J.}~\bibnamefont {Kappler}},\
  and\ \bibinfo {author} {\bibfnamefont {R.~R.}\ \bibnamefont {Netz}},\
  }\bibfield  {title} {\bibinfo {title} {{Barrier crossing in the presence of
  multi-exponential memory functions with unequal friction amplitudes and
  memory times}},\ }\href {https://doi.org/10.1209/0295-5075/131/40004}
  {\bibfield  {journal} {\bibinfo  {journal} {EPL}\ }\textbf {\bibinfo {volume}
  {131}},\ \bibinfo {pages} {40004} (\bibinfo {year} {2020})}\BibitemShut
  {NoStop}%
\bibitem [{\citenamefont {Berezhkovskii}\ and\ \citenamefont
  {Szabo}(2011)}]{Berezhkovskii2011}%
  \BibitemOpen
  \bibfield  {author} {\bibinfo {author} {\bibfnamefont {A.}~\bibnamefont
  {Berezhkovskii}}\ and\ \bibinfo {author} {\bibfnamefont {A.}~\bibnamefont
  {Szabo}},\ }\bibfield  {title} {\bibinfo {title} {{Time scale separation
  leads to position-dependent diffusion along a slow coordinate}},\ }\bibfield
  {journal} {\bibinfo  {journal} {J. Chem. Phys.}\ }\textbf {\bibinfo {volume}
  {135}},\ \href {https://doi.org/10.1063/1.3626215} {10.1063/1.3626215}
  (\bibinfo {year} {2011})\BibitemShut {NoStop}%
\bibitem [{\citenamefont {Hummer}(2004)}]{Hummer2004}%
  \BibitemOpen
  \bibfield  {author} {\bibinfo {author} {\bibfnamefont {G.}~\bibnamefont
  {Hummer}},\ }\bibfield  {title} {\bibinfo {title} {{From transition paths to
  transition states and rate coefficients}},\ }\href
  {https://doi.org/10.1063/1.1630572} {\bibfield  {journal} {\bibinfo
  {journal} {J. Chem. Phys.}\ }\textbf {\bibinfo {volume} {120}},\ \bibinfo
  {pages} {516} (\bibinfo {year} {2004})}\BibitemShut {NoStop}%
\bibitem [{\citenamefont {Hinczewski}\ \emph {et~al.}(2010)\citenamefont
  {Hinczewski}, \citenamefont {von Hansen}, \citenamefont {Dzubiella},\ and\
  \citenamefont {Netz}}]{Hinczewski2010}%
  \BibitemOpen
  \bibfield  {author} {\bibinfo {author} {\bibfnamefont {M.}~\bibnamefont
  {Hinczewski}}, \bibinfo {author} {\bibfnamefont {Y.}~\bibnamefont {von
  Hansen}}, \bibinfo {author} {\bibfnamefont {J.}~\bibnamefont {Dzubiella}},\
  and\ \bibinfo {author} {\bibfnamefont {R.~R.}\ \bibnamefont {Netz}},\
  }\bibfield  {title} {\bibinfo {title} {{How the diffusivity profile reduces
  the arbitrariness of protein folding free energies}},\ }\href
  {https://doi.org/10.1063/1.3442716} {\bibfield  {journal} {\bibinfo
  {journal} {J. Chem. Phys.}\ }\textbf {\bibinfo {volume} {132}},\ \bibinfo
  {pages} {1} (\bibinfo {year} {2010})}\BibitemShut {NoStop}%
\bibitem [{\citenamefont {Straus}\ \emph {et~al.}(1993)\citenamefont {Straus},
  \citenamefont {{Gomez Llorente}},\ and\ \citenamefont {Voth}}]{Straus1993}%
  \BibitemOpen
  \bibfield  {author} {\bibinfo {author} {\bibfnamefont {J.~B.}\ \bibnamefont
  {Straus}}, \bibinfo {author} {\bibfnamefont {J.~M.}\ \bibnamefont {{Gomez
  Llorente}}},\ and\ \bibinfo {author} {\bibfnamefont {G.~A.}\ \bibnamefont
  {Voth}},\ }\bibfield  {title} {\bibinfo {title} {{Manifestations of spatially
  dependent friction in classical activated rate processes}},\ }\href
  {https://doi.org/10.1063/1.465044} {\bibfield  {journal} {\bibinfo  {journal}
  {J. Chem. Phys.}\ }\textbf {\bibinfo {volume} {98}},\ \bibinfo {pages} {4082}
  (\bibinfo {year} {1993})}\BibitemShut {NoStop}%
\bibitem [{\citenamefont {Pollak}\ and\ \citenamefont
  {Berezhkovskii}(1993)}]{Pollak1993}%
  \BibitemOpen
  \bibfield  {author} {\bibinfo {author} {\bibfnamefont {E.}~\bibnamefont
  {Pollak}}\ and\ \bibinfo {author} {\bibfnamefont {A.~M.}\ \bibnamefont
  {Berezhkovskii}},\ }\bibfield  {title} {\bibinfo {title} {{Fokker–Planck
  equation for nonlinear stochastic dynamics in the presence of space and time
  dependent friction}},\ }\href {https://doi.org/10.1063/1.465379} {\bibfield
  {journal} {\bibinfo  {journal} {J. Chem. Phys.}\ }\textbf {\bibinfo {volume}
  {99}},\ \bibinfo {pages} {1344} (\bibinfo {year} {1993})}\BibitemShut
  {NoStop}%
\bibitem [{\citenamefont {Haynes}\ \emph {et~al.}(1994)\citenamefont {Haynes},
  \citenamefont {Voth},\ and\ \citenamefont {Pollak}}]{Haynes1994}%
  \BibitemOpen
  \bibfield  {author} {\bibinfo {author} {\bibfnamefont {G.~R.}\ \bibnamefont
  {Haynes}}, \bibinfo {author} {\bibfnamefont {G.~A.}\ \bibnamefont {Voth}},\
  and\ \bibinfo {author} {\bibfnamefont {E.}~\bibnamefont {Pollak}},\
  }\bibfield  {title} {\bibinfo {title} {{A theory for the activated barrier
  crossing rate constant in systems influenced by space and time dependent
  friction}},\ }\href {https://doi.org/10.1063/1.468274} {\bibfield  {journal}
  {\bibinfo  {journal} {J. Chem. Phys.}\ }\textbf {\bibinfo {volume} {101}},\
  \bibinfo {pages} {7811} (\bibinfo {year} {1994})}\BibitemShut {NoStop}%
\bibitem [{\citenamefont {Singh}\ \emph {et~al.}(1990)\citenamefont {Singh},
  \citenamefont {Krishnan},\ and\ \citenamefont {Robinson}}]{Singh1990}%
  \BibitemOpen
  \bibfield  {author} {\bibinfo {author} {\bibfnamefont {S.}~\bibnamefont
  {Singh}}, \bibinfo {author} {\bibfnamefont {R.}~\bibnamefont {Krishnan}},\
  and\ \bibinfo {author} {\bibfnamefont {G.~W.}\ \bibnamefont {Robinson}},\
  }\bibfield  {title} {\bibinfo {title} {{Theory of activated rate processes
  with space-dependent friction}},\ }\href
  {https://doi.org/10.1016/0009-2614(90)80121-S} {\bibfield  {journal}
  {\bibinfo  {journal} {Chem. Phys. Lett.}\ }\textbf {\bibinfo {volume}
  {175}},\ \bibinfo {pages} {338} (\bibinfo {year} {1990})}\BibitemShut
  {NoStop}%
\bibitem [{\citenamefont {Krishnan}\ \emph {et~al.}(1992)\citenamefont
  {Krishnan}, \citenamefont {Singh},\ and\ \citenamefont
  {Robinson}}]{Krishnan1992}%
  \BibitemOpen
  \bibfield  {author} {\bibinfo {author} {\bibfnamefont {R.}~\bibnamefont
  {Krishnan}}, \bibinfo {author} {\bibfnamefont {S.}~\bibnamefont {Singh}},\
  and\ \bibinfo {author} {\bibfnamefont {G.~W.}\ \bibnamefont {Robinson}},\
  }\bibfield  {title} {\bibinfo {title} {{Space-dependent friction in the
  theory of activated rate processes: The Hamiltonian approach}},\ }\href
  {https://doi.org/10.1063/1.463784} {\bibfield  {journal} {\bibinfo  {journal}
  {J. Chem. Phys.}\ }\textbf {\bibinfo {volume} {97}},\ \bibinfo {pages} {5516}
  (\bibinfo {year} {1992})}\BibitemShut {NoStop}%
\bibitem [{\citenamefont {Singh}\ and\ \citenamefont
  {Robinson}(1994)}]{Singh1994}%
  \BibitemOpen
  \bibfield  {author} {\bibinfo {author} {\bibfnamefont {S.}~\bibnamefont
  {Singh}}\ and\ \bibinfo {author} {\bibfnamefont {G.~W.}\ \bibnamefont
  {Robinson}},\ }\bibfield  {title} {\bibinfo {title} {{Scaling in a model of
  chemical reaction rates with space-dependent friction}},\ }\href
  {https://doi.org/10.1021/j100081a011} {\bibfield  {journal} {\bibinfo
  {journal} {J. Phys. Chem.}\ }\textbf {\bibinfo {volume} {98}},\ \bibinfo
  {pages} {7300} (\bibinfo {year} {1994})}\BibitemShut {NoStop}%
\bibitem [{\citenamefont {Daldrop}\ \emph {et~al.}(2016)\citenamefont
  {Daldrop}, \citenamefont {Kim},\ and\ \citenamefont {Netz}}]{Daldrop2016}%
  \BibitemOpen
  \bibfield  {author} {\bibinfo {author} {\bibfnamefont {J.~O.}\ \bibnamefont
  {Daldrop}}, \bibinfo {author} {\bibfnamefont {W.~K.}\ \bibnamefont {Kim}},\
  and\ \bibinfo {author} {\bibfnamefont {R.~R.}\ \bibnamefont {Netz}},\
  }\bibfield  {title} {\bibinfo {title} {{Transition paths are hot}},\ }\href
  {https://doi.org/10.1209/0295-5075/113/18004} {\bibfield  {journal} {\bibinfo
   {journal} {Europhys. Lett.}\ }\textbf {\bibinfo {volume} {113}},\ \bibinfo
  {pages} {18004} (\bibinfo {year} {2016})}\BibitemShut {NoStop}%
\bibitem [{\citenamefont {Voter}\ and\ \citenamefont {Doll}(1985)}]{Voter1985}%
  \BibitemOpen
  \bibfield  {author} {\bibinfo {author} {\bibfnamefont {A.~F.}\ \bibnamefont
  {Voter}}\ and\ \bibinfo {author} {\bibfnamefont {J.~D.}\ \bibnamefont
  {Doll}},\ }\bibfield  {title} {\bibinfo {title} {{Dynamical corrections to
  transition state theory for multistate systems: Surface self-diffusion in the
  rare-event regime}},\ }\href {https://doi.org/10.1063/1.448739} {\bibfield
  {journal} {\bibinfo  {journal} {J. Chem. Phys.}\ }\textbf {\bibinfo {volume}
  {82}},\ \bibinfo {pages} {80} (\bibinfo {year} {1985})}\BibitemShut {NoStop}%
\bibitem [{\citenamefont {Ayaz}\ \emph {et~al.}(2022)\citenamefont {Ayaz},
  \citenamefont {Dalton},\ and\ \citenamefont {Netz}}]{Ayaz2022}%
  \BibitemOpen
  \bibfield  {author} {\bibinfo {author} {\bibfnamefont {C.}~\bibnamefont
  {Ayaz}}, \bibinfo {author} {\bibfnamefont {B.~A.}\ \bibnamefont {Dalton}},\
  and\ \bibinfo {author} {\bibfnamefont {R.~R.}\ \bibnamefont {Netz}},\
  }\bibfield  {title} {\bibinfo {title} {{Generalized Langevin Equation with a
  Non-Linear Potential of Mean Force and Non-Linear Memory Friction From a
  Hybrid Projection Scheme}},\ }\href {http://arxiv.org/abs/2202.01922}
  {\bibfield  {journal} {\bibinfo  {journal} {Preprint}\ } (\bibinfo {year}
  {2022})},\ \Eprint {https://arxiv.org/abs/2202.01922} {arXiv:2202.01922}
  \BibitemShut {NoStop}%
\bibitem [{\citenamefont {Loos}\ and\ \citenamefont {Klapp}(2020)}]{Loos2020}%
  \BibitemOpen
  \bibfield  {author} {\bibinfo {author} {\bibfnamefont {S.~A.~M.}\
  \bibnamefont {Loos}}\ and\ \bibinfo {author} {\bibfnamefont {S.~H.~L.}\
  \bibnamefont {Klapp}},\ }\bibfield  {title} {\bibinfo {title}
  {{Irreversibility, heat and information flows induced by non-reciprocal
  interactions}},\ }\href {https://doi.org/10.1088/1367-2630/abcc1e} {\bibfield
   {journal} {\bibinfo  {journal} {New J. Phys.}\ }\textbf {\bibinfo {volume}
  {22}},\ \bibinfo {pages} {123051} (\bibinfo {year} {2020})}\BibitemShut
  {NoStop}%
\bibitem [{\citenamefont {Nakajima}(1958)}]{Nakajima1958}%
  \BibitemOpen
  \bibfield  {author} {\bibinfo {author} {\bibfnamefont {S.}~\bibnamefont
  {Nakajima}},\ }\bibfield  {title} {\bibinfo {title} {{On Quantum Theory of
  Transport Phenomena: Steady Diffusion}},\ }\href
  {https://doi.org/10.1143/PTP.20.948} {\bibfield  {journal} {\bibinfo
  {journal} {Prog. Theor. Phys.}\ }\textbf {\bibinfo {volume} {20}},\ \bibinfo
  {pages} {948} (\bibinfo {year} {1958})}\BibitemShut {NoStop}%
\bibitem [{\citenamefont {Ford}\ \emph {et~al.}(1988)\citenamefont {Ford},
  \citenamefont {Lewis},\ and\ \citenamefont {O'Connell}}]{Ford1988}%
  \BibitemOpen
  \bibfield  {author} {\bibinfo {author} {\bibfnamefont {G.~W.}\ \bibnamefont
  {Ford}}, \bibinfo {author} {\bibfnamefont {J.~T.}\ \bibnamefont {Lewis}},\
  and\ \bibinfo {author} {\bibfnamefont {R.~F.}\ \bibnamefont {O'Connell}},\
  }\bibfield  {title} {\bibinfo {title} {{Quantum Langevin equation}},\ }\href
  {https://doi.org/10.1103/PhysRevA.37.4419} {\bibfield  {journal} {\bibinfo
  {journal} {Phys. Rev. A}\ }\textbf {\bibinfo {volume} {37}},\ \bibinfo
  {pages} {4419} (\bibinfo {year} {1988})}\BibitemShut {NoStop}%
\bibitem [{\citenamefont {Carmeli}\ and\ \citenamefont
  {Chandler}(1985)}]{Carmeli1985}%
  \BibitemOpen
  \bibfield  {author} {\bibinfo {author} {\bibfnamefont {B.}~\bibnamefont
  {Carmeli}}\ and\ \bibinfo {author} {\bibfnamefont {D.}~\bibnamefont
  {Chandler}},\ }\bibfield  {title} {\bibinfo {title} {{Effective adiabatic
  approximation for a two level system coupled to a bath}},\ }\href
  {https://doi.org/10.1063/1.448942} {\bibfield  {journal} {\bibinfo  {journal}
  {J. Chem. Phys.}\ }\textbf {\bibinfo {volume} {82}},\ \bibinfo {pages} {3400}
  (\bibinfo {year} {1985})}\BibitemShut {NoStop}%
\bibitem [{\citenamefont {Eyring}(1935)}]{Eyring1935}%
  \BibitemOpen
  \bibfield  {author} {\bibinfo {author} {\bibfnamefont {H.}~\bibnamefont
  {Eyring}},\ }\bibfield  {title} {\bibinfo {title} {{The Activated Complex in
  Chemical Reactions}},\ }\href {https://doi.org/10.1063/1.1749604} {\bibfield
  {journal} {\bibinfo  {journal} {J. Chem. Phys.}\ }\textbf {\bibinfo {volume}
  {3}},\ \bibinfo {pages} {107} (\bibinfo {year} {1935})}\BibitemShut {NoStop}%
\bibitem [{\citenamefont {Arrhenius}(1889)}]{Arrhenius1889}%
  \BibitemOpen
  \bibfield  {author} {\bibinfo {author} {\bibfnamefont {S.}~\bibnamefont
  {Arrhenius}},\ }\bibfield  {title} {\bibinfo {title} {{{\"{U}}ber die
  Reaktionngeschwindigkeit bei des Inversion von Rohrzucker durch
  S{\"{a}}uren}},\ }\href@noop {} {\bibfield  {journal} {\bibinfo  {journal}
  {Zeit. Phys. Chem.}\ }\textbf {\bibinfo {volume} {4}},\ \bibinfo {pages}
  {226} (\bibinfo {year} {1889})}\BibitemShut {NoStop}%
\end{thebibliography}
\onecolumngrid
\clearpage

\end{document}